\documentclass[aps,prl,preprint,tightenlines,superscriptaddress,showpacs,byrevtex]{revtex4}
\usepackage{graphicx} 
\usepackage{dcolumn}  
\usepackage{rotating}

\graphicspath{{ps}}

\newcommand{\Kspipi}{\ensuremath{B^{0} \rightarrow K^{0}_{S} \pi^{+} \pi^{-}}}
\newcommand{\rhozks}{\ensuremath{B^{0} \rightarrow \rho^{0}(770) K^{0}_{S}}}
\newcommand{\fzks}{\ensuremath{B^{0} \rightarrow f_{0}(980) K^{0}_{S}}}
\newcommand{\Ksppim}{\ensuremath{B^{0} \rightarrow K^{*+}(892) \pi^{-}}}

\newcommand{\epem}{\ensuremath{e^{+} e^{-}}}
\newcommand{\Ups}{\ensuremath{\Upsilon (4S)}}
\newcommand{\BBbar}{\ensuremath{B \bar B}}
\newcommand{\BpBm}{\ensuremath{B^{+} B^{-}}}
\newcommand{\BzBzb}{\ensuremath{B^{0} \bar B^{0}}}
\newcommand{\qqbar}{\ensuremath{q \bar q}}

\newcommand{\Mbc}{\ensuremath{M_{\rm bc}}}
\newcommand{\De}{\ensuremath{\Delta E}}

\newcommand{\pipi}{\ensuremath{\pi^{+}\pi^{-}}}
\newcommand{\pipm}{\ensuremath{\pi^{\pm}}}
\newcommand{\pip}{\ensuremath{\pi^{+}}}
\newcommand{\pim}{\ensuremath{\pi^{-}}}

\newcommand{\Kp}{\ensuremath{K^{+}}}

\newcommand{\Ks}{\ensuremath{K^{0}_{S}}}
\newcommand{\Kl}{\ensuremath{K^{0}_{L}}}
\newcommand{\mup}{\ensuremath{\mu^{+}}}
\newcommand{\mum}{\ensuremath{\mu^{-}}}
\newcommand{\Dp}{\ensuremath{D^{+}}}
\newcommand{\Jpsi}{\ensuremath{J/\psi}}
\newcommand{\chic}{\ensuremath{\chi_{c0}}}
\newcommand{\psip}{\ensuremath{\psi(2S)}}
\newcommand{\Ksp}{\ensuremath{K^{*+}(892)}}
\newcommand{\Ksm}{\ensuremath{K^{*-}(892)}}
\newcommand{\Kspm}{\ensuremath{K^{*\pm}(892)}}
\newcommand{\Kstarp}{\ensuremath{K^{*+}_{0}(1430)}}

\newcommand{\Kstarpm}{\ensuremath{K^{*\pm}_{0}(1430)}}
\newcommand{\rhoz}{\ensuremath{\rho^{0}(770)}}
\newcommand{\fz}{\ensuremath{f_{0}(980)}}
\newcommand{\ftwo}{\ensuremath{f_{2}(1270)}}
\newcommand{\fX}{\ensuremath{f_{X}(1300)}}
\newcommand{\Bz}{\ensuremath{B^{0}}}
\newcommand{\Bzb}{\ensuremath{\bar B^{0}}}
\newcommand{\Bp}{\ensuremath{B^{+}}}

\newcommand{\Brec}{\ensuremath{B^{0}_{\rm Rec}}}
\newcommand{\Btag}{\ensuremath{B^{0}_{\rm Tag}}}

\newcommand{\etapks}{\ensuremath{\Bz \rightarrow \eta' \Ks}}
\newcommand{\aonepi}{\ensuremath{\Bz \rightarrow a_{1}(1260)^{+} \pim}}
\newcommand{\aone}{\ensuremath{a_{1}(1260)^{+}}}

\newcommand{\Dt}{\ensuremath{\Delta t}}
\newcommand{\taub}{\ensuremath{\tau_{\Bz}}}
\newcommand{\Dmd}{\ensuremath{\Delta m_{d}}}
\newcommand{\Dw}{\ensuremath{\Delta w}}
\newcommand{\Acp}{\ensuremath{{\cal A}_{CP}}}
\newcommand{\Scp}{\ensuremath{{\cal S}^{\rm eff}_{CP}}}

\newcommand{\phione}{\ensuremath{\phi^{\rm eff}_{1}}}

\begin{document}

\vspace*{\baselineskip}
\resizebox{!}{3cm}{\includegraphics{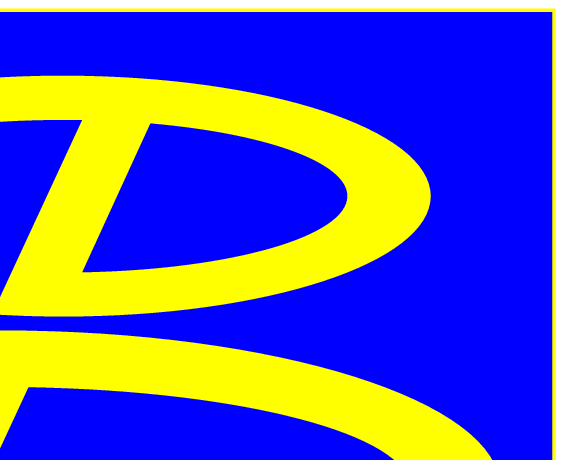}}

\preprint{\vbox{ \hbox{   }
    \hbox{}
    \hbox{}
    \hbox{}
    \hbox{}
    \hbox{}
    \hbox{}
    \hbox{Belle Preprint 2008-29}
    \hbox{KEK Preprint 2008-39}
}}

\title{ \quad\\[1.0cm] Time-dependent Dalitz Plot Measurement of $CP$ Parameters in \Kspipi\ Decays}

\begin{abstract}
  We present a time-dependent Dalitz plot measurement of $CP$ violation parameters in \Kspipi\ decays. These results are obtained from a large data sample that contains $657 \times 10^{6}$ \BBbar\ pairs collected at the \Ups\ resonance with the Belle detector at the KEKB asymmetric-energy \epem\ collider. For the $CP$ violation parameters, we obtain two consistent solutions that describe the data well. The first of these solutions may be preferred by external information from other measurements. There is no evidence for direct $CP$ violation in \rhozks, \fzks\ and \Ksppim, while measurements of mixing-induced $CP$ violation in \rhozks\ and \fzks\ decays are consistent with that of $b \rightarrow c \bar c s$ decays. We also measured the phase difference between $\Bz \rightarrow \Ksp \pim$ and $\Bzb \rightarrow \Ksm \pip$, which may be used to extract $\phi_{3}$.
\end{abstract}

\affiliation{Budker Institute of Nuclear Physics, Novosibirsk}
\affiliation{Chiba University, Chiba}
\affiliation{University of Cincinnati, Cincinnati, Ohio 45221}
\affiliation{The Graduate University for Advanced Studies, Hayama}
\affiliation{Hanyang University, Seoul}
\affiliation{University of Hawaii, Honolulu, Hawaii 96822}
\affiliation{High Energy Accelerator Research Organization (KEK), Tsukuba}
\affiliation{Hiroshima Institute of Technology, Hiroshima}
\affiliation{Institute of High Energy Physics, Chinese Academy of Sciences, Beijing}
\affiliation{Institute of High Energy Physics, Vienna}
\affiliation{Institute of High Energy Physics, Protvino}
\affiliation{Institute for Theoretical and Experimental Physics, Moscow}
\affiliation{J. Stefan Institute, Ljubljana}
\affiliation{Kanagawa University, Yokohama}
\affiliation{Korea University, Seoul}
\affiliation{Kyungpook National University, Taegu}
\affiliation{\'Ecole Polytechnique F\'ed\'erale de Lausanne (EPFL), Lausanne}
\affiliation{Faculty of Mathematics and Physics, University of Ljubljana, Ljubljana}
\affiliation{University of Maribor, Maribor}
\affiliation{University of Melbourne, School of Physics, Victoria 3010}
\affiliation{Nagoya University, Nagoya}
\affiliation{Nara Women's University, Nara}
\affiliation{National Central University, Chung-li}
\affiliation{National United University, Miao Li}
\affiliation{Department of Physics, National Taiwan University, Taipei}
\affiliation{H. Niewodniczanski Institute of Nuclear Physics, Krakow}
\affiliation{Nippon Dental University, Niigata}
\affiliation{Niigata University, Niigata}
\affiliation{University of Nova Gorica, Nova Gorica}
\affiliation{Osaka City University, Osaka}
\affiliation{Osaka University, Osaka}
\affiliation{Panjab University, Chandigarh}
\affiliation{Saga University, Saga}
\affiliation{University of Science and Technology of China, Hefei}
\affiliation{Seoul National University, Seoul}
\affiliation{Sungkyunkwan University, Suwon}
\affiliation{University of Sydney, Sydney, New South Wales}
\affiliation{Toho University, Funabashi}
\affiliation{Tohoku Gakuin University, Tagajo}
\affiliation{Tohoku University, Sendai}
\affiliation{Department of Physics, University of Tokyo, Tokyo}
\affiliation{Tokyo Institute of Technology, Tokyo}
\affiliation{Tokyo Metropolitan University, Tokyo}
\affiliation{Tokyo University of Agriculture and Technology, Tokyo}
\affiliation{IPNAS, Virginia Polytechnic Institute and State University, Blacksburg, Virginia 24061}
\affiliation{Yonsei University, Seoul}
  \author{J.~Dalseno}\affiliation{High Energy Accelerator Research Organization (KEK), Tsukuba} 
  \author{I.~Adachi}\affiliation{High Energy Accelerator Research Organization (KEK), Tsukuba} 
  \author{H.~Aihara}\affiliation{Department of Physics, University of Tokyo, Tokyo} 
  \author{K.~Arinstein}\affiliation{Budker Institute of Nuclear Physics, Novosibirsk} 
  \author{T.~Aushev}\affiliation{\'Ecole Polytechnique F\'ed\'erale de Lausanne (EPFL), Lausanne}\affiliation{Institute for Theoretical and Experimental Physics, Moscow} 
  \author{A.~M.~Bakich}\affiliation{University of Sydney, Sydney, New South Wales} 
  \author{V.~Balagura}\affiliation{Institute for Theoretical and Experimental Physics, Moscow} 
  \author{A.~Bay}\affiliation{\'Ecole Polytechnique F\'ed\'erale de Lausanne (EPFL), Lausanne} 
  \author{V.~Bhardwaj}\affiliation{Panjab University, Chandigarh} 
  \author{U.~Bitenc}\affiliation{J. Stefan Institute, Ljubljana} 
% ...
\collaboration{The Belle Collaboration}


\maketitle


{\renewcommand{\thefootnote}{\fnsymbol{footnote}}}
\setcounter{footnote}{0}

\section{Introduction}

$CP$ violation in the Standard Model (SM) is due to a complex phase in the Cabibbo-Kobayashi-Maskawa (CKM) quark-mixing matrix~\cite{Cabibbo,KM}. At present, mixing-induced $CP$ violation has been clearly observed by the BaBar~\cite{jpsiks_BABAR} and Belle~\cite{jpsiks_Belle} Collaborations in the $b \rightarrow c \bar c s$ induced decay, $\Bz \rightarrow J/\psi \; \Ks$, while many other modes provide additional information on $CP$ violating parameters. Of recent interest is $CP$ violation in $b \rightarrow q \bar q s$ transitions, which proceeds by loop diagrams that may be affected by new particles in various extensions of the SM. Furthermore, the $CP$ asymmetries in $b \rightarrow q \bar q s$ transitions are expected in the SM to be slightly higher than those observed in $b \rightarrow c \bar c s$ transitions~\cite{Theory1,Theory2,Theory3,Theory4,Theory5,Theory6,Theory7,Theory8,Theory9}. However, current experimental measurements~\cite{HFAG} tend to be lower than those for $b \rightarrow c \bar c s$ transitions motivating more precise experimental determinations.

This analysis is concerned with time-dependent $CP$ violation in \Kspipi\ decays including the quasi-two-body modes \rhozks\ and \fzks. Quasi-two-body time-dependent analyses have been performed on \rhozks\ by the BaBar Collaboration~\cite{rho0ks_BABAR} and \fzks\ by both the BaBar and Belle Collaborations~\cite{f0ks_BABAR,f0ks_Belle}. However, a two-body approach to these modes is not ideal due to interference with other resonances as well as non-resonant decays into the same final state. These effects can be taken into account with a time-dependent Dalitz plot fit. In addition, the interference terms in some cases may be sensitive to the cosine of the effective weak phase difference (\phione) in mixing, potentially resolving the two-fold ambiguity in \phione\ characteristic of quasi-two-body analyses. Current direct measurements of $\cos 2\phi_{1}$ from $b \rightarrow c \bar c s$ prefer the solution of $\phi_{1} = 21.5^{\circ}$ over $\phi_{1} = 68.5^{\circ}$~\cite{cos2phi1_BABAR1,cos2phi1_BABAR2,cos2phi1_BABAR3,cos2phi1_Belle1,cos2phi1_Belle2,cos2phi1_Belle3} and similarly, it is important to attempt to resolve this quadratic ambiguity in $b \rightarrow q \bar q s$ modes as well.

Another mode of interest included in \Kspipi\ decays is \Ksppim. Since direct $CP$ violation has been observed in $\Bz \rightarrow \Kp \pim$~\cite{kpi_BABAR, kpi_Belle}, $CP$ violation could also be present in \Ksppim\ decays, though existing measurements are consistent with zero~\cite{ksppim_BABAR}. This measurement may also be related to the $\Acp(K\pi)$ puzzle; the asymmetry in $\Bz \rightarrow \Kp \pim$ is found to be significantly different from that in $B^{\pm} \rightarrow K^{\pm} \pi^{0}$. This unexpected result may indicate the presence of new physics (NP) or poor understanding of strong interaction effects in $B$ decays. A model-independent test for NP is possible via an isospin sum rule with high statistics~\cite{kpi}. Furthermore, it has been proposed that the phase difference between $\Bz \rightarrow \Ksp \pim$ and $\Bzb \rightarrow \Ksm \pip$ can be used to extract $\phi_{3}$~\cite{phi31,phi32,phi33,phi34}. The BaBar collaboration has also performed a time-dependent Dalitz plot analysis on \Kspipi\ decays and released a preliminary result~\cite{kspipi_BaBar}.

\section{Decay Amplitude}
\label{Kspipi Decay Amplitude}
The Dalitz plot variables are defined as the invariant squared masses,
\begin{equation}
  s_{\pm} \equiv (p_{\pm} + p_{0})^{2},
\end{equation}
where $p_{\pm} \, (p_{0})$ are the 4-momenta of the $\pi^{\pm} \, (\Ks)$ respectively. The final combination, $s_{0} \equiv (p_{+} + p_{-})^{2}$ can be obtained from 4-vector conservation,
\begin{equation}
  s_{0} = m^{2}_{\Bz} + 2m^{2}_{\pip} +m^{2}_{\Ks} - s_{+} - s_{-}.
\end{equation}
The differential \Bz\ decay width with respect to the Dalitz plot variables is
\begin{equation}
  d\Gamma(\Kspipi) = \frac{1}{(2\pi)^{3}}\frac{|A|^{2}}{32m^{3}_{\Bz}}ds_{+}ds_{-},
\end{equation}
where $A$ is the Lorentz-invariant amplitude of the decay. In the isobar approximation, the total amplitude of $B^{0}(\Bzb) \rightarrow K^{0}_{S} \pi^{+} \pi^{-}$ is written as the sum of intermediate decay channel amplitudes with the same final state,
\begin{equation}
  A(s_{+}, s_{-}) =\sum_{i} a^{\prime}_{i}F_{i}(s_{+}, s_{-}), \;\;\;\; \bar A(s_{-}, s_{+}) =\sum_{i} \bar a^{\prime}_{i} \bar F_{i}(s_{-}, s_{+}),
\end{equation}
where $a^{\prime}_{i}\equiv |a^{\prime}_{i}|e^{ib^{\prime}_{i}}$ are complex coefficients describing the relative magnitudes and phases between the decay channels, which also carry the weak phase dependence. The amplitudes, $F_{i}(s_{+}, s_{-})$ contain only strong dynamics and thus, $F_{i}(s_{+}, s_{-}) = \bar F_{i}(s_{-}, s_{+})$. They can be expanded in terms of invariant mass and angular distribution probabilities,
\begin{equation}
  F^{L}_{i}(s_{+}, s_{-}) \equiv X^{L}_{i}(\vec p^{\;*}) \times X^{L}_{i}(\vec q) \times Z^{L}_{i}(\vec p, \vec q) \times R_{i}(s_{+}, s_{-}),
\end{equation}
where $\vec p^{\;*}$ is the momentum of the bachelor particle in the \Bz\ rest frame, $\vec p$ and $\vec q$ are the momenta of the bachelor particle and one of the resonance daughters in the resonance's rest frame respectively, $L$ is the orbital angular momentum between the resonance and the bachelor particle while $X^{L}_{i}$ are the Blatt-Weisskopf barrier factors~\cite{BWBF}. As these factors are unknown in general and the Blatt-Weisskopf forms do not improve this model, they are taken to be unity with the uncertainty in this choice being treated in the systematic errors. The angular distribution, $Z^{L}_{i}(\vec p, \vec q)$ depends on $L$,
\begin{eqnarray}
    Z^{0}_{i}(\vec p, \vec q) \!\!\!&=&\!\!\! 1, \nonumber \\
    Z^{1}_{i}(\vec p, \vec q) \!\!\!&=&\!\!\! -4\vec p \cdot \vec q ,\nonumber \\
    Z^{2}_{i}(\vec p, \vec q) \!\!\!&=&\!\!\! \frac{8}{3}[3(\vec p \cdot \vec q)^{2} - (|\vec p||\vec q|)^{2}].
\end{eqnarray}
The mass shapes are denoted as $R_{i}(s_{+}, s_{-})$, which differ depending on the decay channel. We utilize the Relativistic Breit-Wigner (RBW)~\cite{PDG}, Gounaris-Sakurai (GS)~\cite{GS} and Flatt\'{e}~\cite{Flatte} line shapes. Table~\ref{tab_sigmod} summarizes the components considered in the signal model, which was motivated by the previous Belle Dalitz plot analysis of \Kspipi~\cite{kspipi_Belle1}. The \fX\ resonance of unknown spin that appears in the table was first described in Ref.~\cite{kspipi_Belle1} and is assumed to be a scalar.
\begin{table}[htb]
  \caption{Summary of the resonances considered in the signal model. All fixed parameters are taken from Ref.~\cite{PDG} with the exception of those for the \fz\ and \fX, which are taken from Refs.~\cite{fz}~and~\cite{kspipi_Belle1}, respectively.}
  \label{tab_sigmod}
  \begin{tabular}
    {@{\hspace{0.5cm}}c@{\hspace{0.5cm}}  @{\hspace{0.5cm}}c@{\hspace{0.5cm}}  @{\hspace{0.5cm}}c@{\hspace{0.5cm}}}
    \hline \hline
    Resonance & Fixed Parameters $({\rm MeV}/c^{2})$ & Form Factor, $R_{i}(s_{+}, s_{-})$\\
    \hline
    \Kspm & ${\rm m} = 891.7 \pm 0.3$ & RBW\\
    & $\Gamma = 50.8 \pm 0.9$ &\\
    \Kstarpm & ${\rm m} = 1414 \pm 6$ & RBW\\
    & $\Gamma = 290 \pm 21$ &\\
    \rhoz & ${\rm m} = 775.5 \pm 0.3$ & GS\\
    & $\Gamma = 146.2 \pm 0.7$ &\\
    \fz & ${\rm m} = 965 \pm 10$ & Flatt\'{e}\\
    & $g_{\pi} = 0.165 \pm 0.018$ ${\rm GeV}^{2}/c^{4}$&\\
    & $g_{K} = (4.21 \pm 0.09)g_{\pi}$ ${\rm GeV}^{2}/c^{4}$&\\
    \ftwo & ${\rm m} = 1275 \pm 1$ & RBW\\
    & $\Gamma = 185 \pm 3$ &\\
    \fX & ${\rm m} = 1449 \pm 13$ & RBW\\
    & $\Gamma = 126 \pm 25$ &\\
    $(\Ks \pip)_{\rm NR} \pim$ & & $e^{-\alpha s^{+}}$\\
    $(\Ks \pim)_{\rm NR} \pip$ & & $e^{-\alpha s^{-}}$\\
    $(\pip \pim)_{\rm NR} \Ks$ & & $e^{-\alpha s^{0}}$\\
    \hline \hline
  \end{tabular}
\end{table}

\section{Time-dependence}
The decay of the \Ups\ produces a \BBbar\ pair of which one (\Brec) may be reconstructed as \Kspipi\ while the other (\Btag) may reveal its flavor. The proper time interval between the \Brec\ and \Btag\ mesons that decay at times, $t_{\rm Rec}$ and $t_{\rm Tag}$, respectively, is defined as $\Dt \equiv t_{\rm Rec} - t_{\rm Tag}$. For coherent \BBbar\ production in a \Ups\ decay, the time-dependent decay rate when \Btag\ possesses flavor, $q$ (\Bz: $q=+1$, \Bzb: $q=-1$), is given by
\begin{eqnarray}
  |A(\Dt, q)|^{2} = \frac{e^{-|\Dt|/\taub}}{4\taub} &\biggl[& (|A|^{2} + |\bar A|^{2}) - q(|A|^{2} - |\bar A|^{2}) \cos \Dmd \Dt \nonumber \\
    &&\!\!\!\!\!\! + 2q \Im (\bar A A^{*}) \sin \Dmd \Dt \biggr],
\end{eqnarray}
where \taub\ is the \Bz\ lifetime and \Dmd\ is the mass difference between the two mass eigenstates of the neutral $B$ meson. This assumes no $CP$ violation in mixing, $|q/p|=1$, and that the total decay rate difference between the two mass eigenstates is negligible. The amplitudes, $A$, were defined previously; we choose a convention where the \Bz-\Bzb\ mixing phase of $q/p$ is absorbed into the \Bzb\ decay amplitude, $\bar a^{\prime}_{i}$.

These complex coefficients can be redefined in a way that depends on the decay amplitude,
\begin{equation}
  a^{\prime}_{i} \equiv a_{i}(1 + c_{i})e^{i(b_{i} + d_{i})}
\end{equation}
for $A$ and,
\begin{equation}
  \bar a^{\prime}_{i} \equiv a_{i}(1 - c_{i})e^{i(b_{i} - d_{i})}
\end{equation}
for $\bar A$, and thus a resonance, $i$, has a direct $CP$ violation asymmetry given by
\begin{equation}
  \Acp(i) \equiv \frac{|\bar a^{\prime}_{i}|^{2}-|a^{\prime}_{i}|^{2}}{|\bar a^{\prime}_{i}|^{2}+|a^{\prime}_{i}|^{2}} =  \frac{-2c_{i}}{1+c_{i}^{2}}.
\end{equation}
For a $CP$ eigenstate, the CKM angle, $\phione(i)$, is reduced to a fit parameter,
\begin{equation}
  \phione(i) \equiv \frac{\arg(a^{\prime}_{i} \bar a^{\prime *}_{i})}{2} = d_{i},
\end{equation}
and its effective mixing-induced $CP$ violation asymmetry is calculated as
\begin{equation}
   -\eta_{i}\Scp(i) \equiv \frac{-2\Im(\bar a^{\prime}_{i}a^{\prime *}_{i})}{|a^{\prime}_{i}|^{2} + |\bar a^{\prime}_{i}|^{2}} = \frac{1-c^{2}_{i}}{1+c^{2}_{i}}\sin 2 \phione(i),
\end{equation}
where $\eta_{i}$ is the $CP$ eigenvalue of the final state. Note that $\Acp(i)$ and $\Scp(i)$ are restricted by these definitions to lie in the physical region. For flavor-specific states the phase difference is calculated as
\begin{equation}
  \Delta \phi(i) \equiv \arg(a^{\prime}_{i} \bar a^{\prime *}_{i}) = 2d_{i}.
\end{equation}

\section{Data Set And Belle Detector}
This time-dependent Dalitz plot measurement of $CP$ violating parameters in \Kspipi\ is based on a data sample that contains $657 \times 10^6$ \BBbar\ pairs collected with the Belle detector at the KEKB asymmetric-energy \epem\ ($3.5$ on $8~{\rm GeV}$) collider~\cite{KEKB}. Operating with a peak luminosity that exceeds $1.7\times 10^{34}~{\rm cm}^{-2}{\rm s}^{-1}$, the collider produces the \Ups\ resonance ($\sqrt{s}=10.58$~GeV) with a Lorentz boost of $\beta\gamma=0.425$, opposite to the positron beam direction, $z$, which usually decays into a \BBbar\ pair.

The Belle detector is a large-solid-angle magnetic
spectrometer that
consists of a silicon vertex detector (SVD),
a 50-layer central drift chamber (CDC), an array of
aerogel threshold Cherenkov counters (ACC), 
a barrel-like arrangement of time-of-flight
scintillation counters (TOF), and an electromagnetic calorimeter (ECL)
comprised of CsI (Tl) crystals located inside 
a superconducting solenoid coil that provides a 1.5~T
magnetic field.  An iron flux-return located outside of
the coil is instrumented to detect \Kl\ mesons and to identify
muons (KLM).  The detector
is described in detail elsewhere~\cite{Belle}.
Two inner detector configurations were used. A 2.0 cm beampipe
and a 3-layer silicon vertex detector (SVD1) was used for the first sample
of 152 $\times 10^6 B\bar{B}$ pairs, while a 1.5 cm beampipe, a 4-layer
silicon detector (SVD2) and a small-cell inner drift chamber was used to record
the remaining 505 $\times 10^6 B\bar{B}$ pairs~\cite{SVD2}. We use a GEANT-based Monte Carlo (MC) simulation to model the response of the detector and determine its acceptance~\cite{GEANT}.

\section{Event Selection}
We reconstruct $B$ candidates from a \Ks\ candidate and a pair of oppositely-charged tracks. Charged tracks satisfy loose criteria on their impact parameters relative to the interaction point (IP), $dr < 0.4 \; {\rm cm}$ and $|dz| < 5.0 \; {\rm cm}$, where $r$ is the radial coordinate of the Belle detector. With information obtained from the CDC, ACC and TOF, particle identification (PID) is determined with the likelihood ratio, ${\cal L}_{i}/({\cal L}_{i} + {\cal L}_{j})$. Here, ${\cal L}_{i}$ (${\cal L}_{j}$) is the likelihood that the particle is of type $i$ ($j$). To suppress background from particle misidentification, vetoes are applied on particles consistent with kaon, electron or proton hypotheses. The transverse momenta of charged tracks are required to be greater than $100 \; {\rm MeV}$ and the additional SVD requirements of two $z$ hits and one $r-\phi$ hit~\cite{ResFunc} are imposed so that a good quality vertex of the reconstructed $B$ candidate can be determined. We only consider $\Ks \rightarrow \pip \pim$ candidates with vertices that are displaced from the IP and that lie in the mass window, $|m(\pip \pim) - m(\Ks)| < 15 \; {\rm MeV}/c^{2}$.

Reconstructed $B$ candidates are described with two kinematic variables: the beam-constrained mass, $\Mbc \equiv \sqrt{(E^{\rm CMS}_{\rm beam})^{2} - (p^{\rm CMS}_{B})^{2}}$ and the energy difference, $\De \equiv E^{\rm CMS}_{B} - E^{\rm CMS}_{\rm beam}$ where $E^{\rm CMS}_{\rm beam}$ is the beam energy and $E^{\rm CMS}_{B}$ ($p^{\rm CMS}_{B}$) is the energy (momentum) of the $B$ meson all evaluated in the center-of-mass system (CMS). The signal region is defined as, $5.27 \; {\rm GeV}/c^{2} < \Mbc < 5.29 \; {\rm GeV}/c^{2}$ and $-0.04 \; {\rm GeV} < \De < 0.04 \; {\rm GeV}$. The dominant background in the reconstruction of \Brec\ is from continuum ($\epem \rightarrow \qqbar$) events. Since their topology tends to be jet-like in contrast to the spherical \BBbar\ decay, continuum can be suppressed with a Fisher discriminant based on modified Fox-Wolfram moments~\cite{KSFW}. This discriminant is combined with the polar angle of the $B$ candidate in the CMS, $\cos \theta_{B}$, which follows a $1-\cos^{2} \theta_{B}$ distribution for \BBbar\ events while being flat for continuum. A requirement that rejects 91\% of the continuum background while retaining 73\% of the signal events is applied. The second largest background comes from charm decays of the $B$ meson and peaks in the signal region. We apply the charm vetoes summarized in Table~\ref{tab_veto}. The fraction of events having more than one reconstructed candidate in regions dominanted by particular quasi-two-body modes is $28\%$ for \rhozks, $25\%$ for \fzks\ and $21\%$ for \Ksppim. Selecting the $B$ candidate having an \Mbc\ value closest to the nominal $B$ meson mass, the fraction of misreconstructed events is $3\%$ for \rhozks, $2\%$ for \fzks\ and $2\%$ for \Ksppim. This criteria introduces negligible bias into the \De\ distribution. From this best candidate, the Dalitz plot coordinates in the signal region, $s_{\pm}$, are calculated after a mass-constrained fit of the $B$ meson to ensure all the events are within the Dalitz plot boundaries.
\begin{table}[htb]
  \caption{Summary of charm vetoes. The subscript in the region vetoed indicates that an alternate mass hypothesis has been applied to the pion candidates used to calculate the invariant mass term.}
  \label{tab_veto}
  \begin{tabular}
    {@{\hspace{0.5cm}}c@{\hspace{0.5cm}}  @{\hspace{0.5cm}}c@{\hspace{0.5cm}}}
    \hline \hline
    Region vetoed & Mode vetoed\\
    \hline
    $|m(\Ks \pipm) - m(\Dp)| < 100 \; {\rm MeV}/c^{2}$ & $\Bzb \rightarrow D^{+} [\Ks \pip] \; \pim$\\
    $|m(\Ks \pipm)_{K} - m(\Dp)| < 15 \; {\rm MeV}/c^{2}$ & $\Bzb \rightarrow D^{+} [\Ks \Kp] \;\pim$\\
    $|m(\pip \pim)_{\mu} - m(\Jpsi)| < 70 \; {\rm MeV}/c^{2}$ & $\Bzb \rightarrow \Jpsi [\mup \mum] \; \Ks$\\
    $|m(\pip \pim) - m(\chic)| < 70 \; {\rm MeV}/c^{2}$ & $\Bzb \rightarrow \chic [\pip \pim] \; \Ks$\\
    $|m(\pip \pim)_{\mu} - m(\psip)| < 50 \; {\rm MeV}/c^{2}$ & $\Bzb \rightarrow \psip [\mup \mum] \; \Ks$\\
    \hline \hline
  \end{tabular}
\end{table}

Since the \Brec\ and \Btag\ mesons are approximately at rest in the \Ups\ CMS, the difference in decay time between the two $B$ mesons, $\Delta t$, can be determined from the displacement in $z$ between the final state decay vertices,
\begin{equation}
  \Dt \simeq \frac{(z_{\rm Rec} - z_{\rm Tag})}{\beta \gamma c} \equiv \frac{\Delta z}{\beta \gamma c}.
\end{equation}
The vertex of reconstructed $B$ candidates is determined from the charged daughters using the known IP. The IP profile is smeared in the plane perpendicular to $z$ to account for the finite flight length of the $B$ meson in this plane. To obtain the \Dt\ distribution, we reconstruct the tag-side vertex from the tracks not used to reconstruct \Brec~\cite{ResFunc} and employ the flavor tagging routine described in Ref.~\cite{Tagging}. The tagging information is represented by two parameters, the \Btag\ flavor, $q$ and $r$. The parameter $r$ is an event-by-event, MC determined flavor-tagging dilution factor that ranges from $r = 0$ for no flavor discrimination to $r = 1$ for unambiguous flavor assignment. The total effective tagging efficiency is determined to be $0.29 \pm 0.01$.

\section{Signal Yield}
\begin{figure}
  \centering
  \includegraphics[height=200pt,width=!]{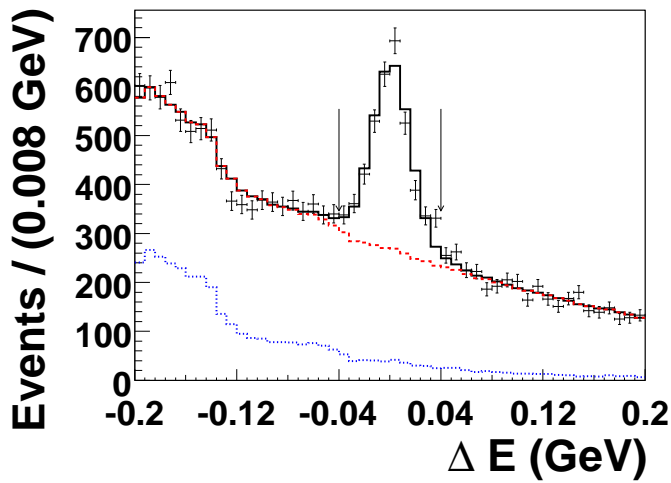}
  \caption{Fit result for \Kspipi signal yield. The solid curve shows the fit result, the dotted curve represents \BBbar\ and the dashed curve is the sum of the \qqbar\ and \BBbar\ components.}
  \label{fig_ede}
\end{figure}

The signal yield is extracted with an extended one-dimensional unbinned maximum likelihood fit to \De\ with the requirement, $5.27 \; {\rm GeV}/c^{2} < \Mbc < 5.29 \; {\rm GeV}/c^{2}$. The signal probability density function (PDF), ${\cal P}_{\rm Sig}$, is modeled with a sum of two Gaussians, where the fraction and shape parameters of the tail Gaussian are fixed from MC relative to the main Gaussian. The \qqbar\ component, ${\cal P}_{q \bar q}$, is modeled with a first-order polynomial whose slope is allowed to float in the fit while the \BBbar\ component, ${\cal P}_{\BBbar}$ is described by a histogram determined from MC. The total likelihood for $N$ events becomes
\begin{equation}
  {\cal L} = \frac{e^{-(N_{\rm Sig}+N_{\rm Bkg})}}{N!} \prod^{N}_{i=1}  N_{\rm Sig}{\cal P}_{\rm Sig}(\De_{i}) + N_{\rm Bkg}[f_{q \bar q}{\cal P}_{q \bar q}(\De_{i}) + (1-f_{q \bar q}) {\cal P}_{\BBbar}(\De_{i})],
\end{equation}
where the \qqbar\ fraction, $f_{q \bar q}$, is a free parameter. A fit to data yields $N_{\rm Sig} = 1944 \pm 98$ events with the fit result shown in Fig.~\ref{fig_ede}. In the signal region, there are a total of $4547$ \Kspipi\ candidates that are used to perform the time-dependent Dalitz plot analysis. The average signal purity in this region is $0.41 \pm 0.02$ and the contribution of continuum to the total background is found to be $0.89 \pm 0.02$.

\section{Square Dalitz Plot}
In \Kspipi\ decays, both signal and continuum events tend to concentrate near the Dalitz plot kinematic boundaries making large variations in these small areas difficult to describe if non-parametric shapes such as histograms are used to describe the continuum background. In order to alleviate this problem, a transformation that expands these regions,
\begin{equation}
ds_{+}ds_{-} \rightarrow |{\rm det} J| dm' d\theta',
\end{equation}
is introduced to define the Square Dalitz Plot~\cite{SDP}. Its coordinates are
\begin{equation}
  m' \equiv \frac{1}{\pi}\arccos \biggl( 2\frac{m_{0} - m^{\rm min}_{0}}{m^{\rm max}_{0} - m^{\rm min}_{0}}-1 \biggl), \;\;\;\; \theta' \equiv \frac{1}{\pi} \theta_{0},
\end{equation}
where $m_{0}$ is the \pipi\ invariant mass, $m^{\rm min}_{0} = 2m_{\pip}$, $m^{\rm max}_{0} = m_{\Bz} - m_{\Ks}$ are the kinematic limits and $\theta_{0}$ is the angle between the \pim\ and \Ks\ in the \pipi\ frame. The determinant of the Jacobian of this transformation is
\begin{equation}
  |{\rm det} J| = 4|\vec p_{-}||\vec p_{0}|m_{0}\frac{\pi^{2}(m^{\rm max}_{0} - m^{\rm min}_{0})}{2} \sin \pi m'  \sin \pi \theta',
\end{equation}
with $|\vec p_{-}| = \sqrt{E^{2}_{-}-m^{2}_{\pim}}$ and $|\vec p_{0}| = \sqrt{E^{2}_{0}-m^{2}_{\Ks}}$ evaluated in the \pipi\ frame.

\section{Signal PDF}
The signal PDF is given by
\begin{eqnarray}
  {\cal P}_{\rm True}(m', \theta', \Dt, q) = \epsilon(m', \theta') |{\rm det} \; J| \frac{e^{-|\Dt|/\taub}} {4\taub} &\biggl[& (1-q\Dw)(|A|^{2} + |\bar A|^{2} ) \nonumber \\
    && - q (1-2w) (|A|^{2} - |\bar A|^{2}) \cos \Dmd \Dt \nonumber \\
    && + 2q(1-2w) \Im (\bar A A^{*}) \sin \Dmd \Dt \biggr] \otimes R_{\BzBzb}, \nonumber \\
\end{eqnarray}
which accounts for $CP$ dilution from the probability of incorrect flavor tagging, $w$ and the wrong tag difference between \Bz\ and \Bzb, \Dw, both of which are determined from flavor specific control samples~\cite{Tagging}. This PDF is convolved with the \Dt\ resolution function, $R_{\BzBzb}$, described in Ref.~\cite{ResFunc}. On the other hand, detector resolution is ignored in the Dalitz plot because the widths of the dominant resonances are large compared to the mass resolution. For example, the detector resolution in the region of the narrowest resonance, the \fz, is typically $4$ MeV, while most estimates of the \fz\ width tend to be at least an order of magnitude larger~\cite{PDG}. The relative signal detection efficiency variation across the Dalitz plane due to detector acceptance, $\epsilon(m', \theta')$, is also considered and is taken from MC. Misreconstructed events usually occur around the corners of the Dalitz plot with a slow pion from the tag side. The fraction of misreconstruction can be as high as $10\%$ in these regions. As misreconstruction is not accounted for in the signal model, its effect will be included in the systematic errors.

\section{Continuum PDF}
The PDF for the time-dependent Dalitz plot for the \qqbar\ continuum is determined from a narrow sideband, $5.250 \; {\rm GeV}/c^{2} < \Mbc < 5.265 \; {\rm GeV}/c^{2}$ and $-0.1 \; {\rm GeV} < \De < 0.1 \; {\rm GeV}$, which ensures that the resonances present in continuum are not shifted significantly as the Dalitz variables are scaled to originate from $B$ decays. There is a flavor tag asymmetry in $\theta'$ due to the jet-like topology of continuum; as a high momentum \pip(\pim) in \Brec\ is correlated with a high momentum \pim(\pip) on the tag side. To account for this, we fit the Dalitz plot asymmetry with an empirical two-dimensional PDF,
\begin{equation}
  {\cal A}_{\qqbar}(m', \theta'; B, C) = [m'(B+Cm')](\theta'-0.5).
\end{equation}
The Dalitz plot component of the \qqbar\ PDF will be modeled with a two-dimensional histogram, $H_{\qqbar}(m', \theta')$, including a term accounting for the flavor tag asymmetry with parameters fixed from the two-dimensional fit to sideband. To reduce the effects of statistical fluctuations with limited data, the symmetry condition, $H_{\qqbar}(m', \theta') = H_{\qqbar}(m', 1-\theta')$ is imposed.

The \Dt\ distribution is modeled with a lifetime and prompt component,
\begin{equation}
  P_{\qqbar}(\Dt) \equiv (1 - f_{\delta}) \frac{e^{-|\Dt|/\tau_{\qqbar}}}{2\tau_{\qqbar}} + f_{\delta} \; \delta( \Dt - \mu_{\delta}),
\end{equation}
convolved with a double Gaussian,
\begin{equation}
  R_{\qqbar}(\Dt) = (1-f_{\rm tail})G(\Dt; \mu_{\rm mean}, S_{\rm main}\sigma) + f_{\rm tail}G(\Dt; \mu_{\rm mean}, S_{\rm main}S_{\rm tail}),
\end{equation}
which uses the event-dependent \Dt\ error constructed from the vertex resolution, $\sigma \equiv (\sqrt{\sigma^{2}_{\rm Rec}+\sigma^{2}_{\rm Tag}})/\beta \gamma c$ as a scale factor. The total \qqbar\ PDF is given by
\begin{equation}
  {\cal P}_{\qqbar}(m', \theta', \Dt, q) = \frac{1 + q{\cal A}_{\qqbar}(m', \theta')}{2} H_{\qqbar}(m', \theta') P_{\qqbar}(\Dt).
\end{equation}
As a check of this PDF, we perform fits in various Dalitz plot regions to extract the \Dt\ shape, however no correlation between \Dt\ and the Dalitz plot coordinates was found.

\section{Charged $B$ PDF}
Like continuum, \BpBm\ events also exhibit a flavor tag asymmetry. Using a generic MC sample that contains charm and rare $B$ decays, a similar procedure is performed to model the asymmetry in the Dalitz plane. The Dalitz component of the \BpBm\ PDF is modeled with a two-dimensional histogram symmetrized as $H_{\BpBm}(m', \theta') = H_{\BpBm}(m', 1-\theta')$.

Since charged $B$ events are misreconstructed by borrowing a particle from the tag side, the average \Dt\ lifetime tends to be smaller and should be taken into account. The \Dt\ distribution is modeled with an exponential PDF with effective lifetimes determined from MC, convolved with the \Dt\ resolution function for \Bp, $R_{\BpBm}$,
\begin{equation}
  P_{\BpBm}(\Dt) \equiv \frac{e^{-|\Dt|/\tau_{\rm eff}} }{2\tau_{\rm eff}} \otimes R_{\BpBm}.
\end{equation}
The \BpBm\ time-dependent Dalitz plot PDF becomes
\begin{equation}
  {\cal P}_{\BpBm}(m', \theta', \Dt, q) = \frac{1 + q{\cal A}_{\BpBm}(m', \theta')}{2} H_{\BpBm}(m', \theta') P_{\BpBm}(\Dt).
\end{equation}

\section{Neutral $B$ PDF}
As in the charged $B$ case, this PDF is also fixed from MC. However, no correlation between flavor and the Dalitz coordinate is found. A two-dimensional histogram is chosen to represent the Dalitz distribution that again incorporates the symmetry condition, $H_{\BzBzb}(m', \theta') = H_{\BzBzb}(m', 1-\theta')$.

Lifetime fits to the \Dt\ distribution are also consistent with the nominal lifetime value, therefore the current world-average \Bz\ lifetime will be used as the default value in all fits. As neutral \BBbar\ pairs mix, a time-dependent $CP$ violating PDF is used to model \Dt\ and $q$,
\begin{eqnarray}
  P_{\BzBzb}(\Dt, q) && \!\!\!\! \equiv \nonumber \\
  && \!\!\!\!\!\!\!\!\!\!\!\!\!\!\!\!\!\! \frac{e^{-|\Dt|/\tau_{\Bz}} }{4 \tau_{\Bz}} \biggl[ 1 - q \Delta w + q(1 - 2w) \biggl( {\cal A}_{\BzBzb} \cos \Dmd \Dt + {\cal S}_{\BzBzb} \sin \Dmd \Dt \biggr) \biggr] \otimes R_{\BzBzb}, \nonumber \\
  \label{eq_qqbar_tcpv_pdf}
\end{eqnarray}
with ${\cal A}_{\BzBzb} = {\cal S}_{\BzBzb} = 0$. The total \BzBzb\ PDF is,
\begin{equation}
  {\cal P}_{\BzBzb}(m', \theta', \Dt, q) = H_{\BzBzb}(m', \theta') P_{\Bz \Bzb}(\Dt, q).
\end{equation}
Significant \BzBzb\ backgrounds with measured $CP$ parameters are treated separately using Eq.~\ref{eq_qqbar_tcpv_pdf}. There are $\sim 20$ \etapks\ and $\sim 4$ \aonepi\ events expected in the data sample. In these decay channels no flavor tag asymmetry in MC is found and lifetime measurements are consistent with the nominal value. The $CP$ parameters for \etapks\ and \aonepi\ are taken from Ref.~\cite{PDG}.

\section{Outlier PDF}
To account for events with \Dt\ values not yet described by either signal or background PDFs, an outlier PDF is introduced,
\begin{equation}
  {\cal P}_{\rm Out}(m', \theta', \Dt, q) = \frac{1}{2} H(m', \theta') G(\Dt; 0, \sigma_{\rm Out}),
\end{equation}
where the width of the Gaussian, $\sigma_{\rm Out}$, is determined in Ref.~\cite{ResFunc} and $H(m', \theta')$ is the Dalitz plot histogram of data.

\section{Time-dependent Dalitz Plot PDF}
The full time-dependent Dalitz plot PDF is given by
\begin{eqnarray}
  && \!\!\!\!\!\!\!\!\!\!\!\!\!\!\!\!\!\!\!\! {\cal P}(m', \theta', \Dt, q; \De, r) = \nonumber \\
  && (1-f_{\rm Out})\{f_{\rm Sig}(\De, r){\cal P}_{\rm Sig}(m', \theta', \Dt, q) + \nonumber \\
  && \;\;\;\;\;\;\;\;\;\;\;\;\;\;\;\;\; [1-f_{\rm Sig}(\De, r)][f_{\qqbar}(\De){\cal P}_{\qqbar}(m', \theta', \Dt, q) + \nonumber \\
  && \;\;\;\;\;\;\;\;\;\;\;\;\;\;\;\;\;\;\;\;\;\;\;\;\;\;\;\;\;\;\;\;\;\;\;\;\;\;\;\;\;\;\;\; \{1-f_{\qqbar}(\De)\}(f_{\BpBm}{\cal P}_{\BpBm}(m', \theta', \Dt, q) + \nonumber \\
  && \;\;\;\;\;\;\;\;\;\;\;\;\;\;\;\;\;\;\;\;\;\;\;\;\;\;\;\;\;\;\;\;\;\;\;\;\;\;\;\;\;\;\;\;\;\;\;\;\;\;\;\;\;\;\;\;\;\;\;\;\;\;\;\;\;\;\;\; f_{\BzBzb}{\cal P}_{\BzBzb}(m', \theta', \Dt, q) + \nonumber \\
  && \;\;\;\;\;\;\;\;\;\;\;\;\;\;\;\;\;\;\;\;\;\;\;\;\;\;\;\;\;\;\;\;\;\;\;\;\;\;\;\;\;\;\;\;\;\;\;\;\;\;\;\;\;\;\;\;\;\;\;\;\;\;\;\;\;\;\;\; f_{\eta' \Ks}{\cal P}_{\eta' \Ks}(m', \theta', \Dt, q) + \nonumber \\
  && \;\;\;\;\;\;\;\;\;\;\;\;\;\;\;\;\;\;\;\;\;\;\;\;\;\;\;\;\;\;\;\;\;\;\;\;\;\;\;\;\;\;\;\;\;\;\;\;\;\;\;\;\;\;\;\;\;\;\;\;\;\;\;\;\;\;\;\; (1-f_{\BpBm}-f_{\BzBzb}-f_{\eta' \Ks}) \nonumber \\
  &&\;\;\;\;\;\;\;\;\;\;\;\;\;\;\;\;\;\;\;\;\;\;\;\;\;\;\;\;\;\;\;\;\;\;\;\;\;\;\;\;\;\;\;\;\;\;\;\;\;\;\;\;\;\;\;\;\;\;\;\;\;\;\;\;\;\;\;\; \times {\cal P}_{\aone \pim}(m', \theta', \Dt, q) )]\} + \nonumber \\
  && f_{\rm Out}{\cal P}_{\rm Out}(m', \theta', \Dt, q),
  \label{eq_dpdt_tcpv_pdf}
\end{eqnarray}
where $f_{\rm Sig}(\De, r)$ is the event-dependent signal probability,
\begin{equation}
  f_{\rm Sig}(\De, r) = \frac{p_{\rm Sig}(r){\cal P}_{\rm Sig}(\De)}{p_{\rm Sig}(r){\cal P}_{\rm Sig}(\De) + [1-p_{\rm Sig}(r)][p_{\qqbar}{\cal P}_{\qqbar}(\De)+(1-p_{\qqbar}){\cal P}_{\BBbar}(\De)]}.
\end{equation}
The $r$-bin dependent purity, $p_{\rm Sig}(r)$, is calculated in the signal region and is included to increase signal sensitivity and $p_{\qqbar} \equiv N_{\qqbar}/(N_{\qqbar}+N_{\BBbar})$ is the \qqbar\ contribution to the total background in the signal region. The event-dependent \qqbar\ probability is calculated as
\begin{equation}
  f_{\qqbar}(\De) = \frac{p_{\qqbar}{\cal P}_{\qqbar}(\De)}{p_{\qqbar}{\cal P}_{\qqbar}(\De) + (1-p_{\qqbar}){\cal P}_{\BBbar}(\De)},
\end{equation}
and $f_{\BpBm}$, $f_{\BzBzb}$, $f_{\eta' \Ks}$ are constants defined in the signal region as
\begin{equation}
    f_{i} = \frac{N_{i}}{N_{\BpBm} + N_{\BzBzb} + N_{\eta' \Ks} + N_{\aone \pim}},
\end{equation}
where $N_{i}$ are the expected number of events in background category $i$.

As there is only sensitivity to the relative amplitudes and phases between decay modes, we fix $a_{\Ksp} = 1$ and $b_{\Ksp} = 0$. In addition, the \ftwo, \fX\ and non-resonant components share common $CP$ parameters giving a total of $27$ free parameters.

\section{Fit Result}
We perform a time-dependent Dalitz plot fit to data and find four solutions with consistent $CP$ parameters given in Table~\ref{tab_dtcpv}. These were obtained by performing numerous fits with randomly generated initial values for the free parameters. For each resonance, $i$, a relative fraction can be calculated as
\begin{equation}
  f_{i} = \frac{(|a^{\prime}_{i}|^{2} + |\bar a^{\prime}_{i}|^{2}) \int F_{i}(s_{+}, s_{-}) F^{*}_{i}(s_{+}, s_{-}) ds_{+} ds_{-}}{\int  (|A|^{2} + |\bar A|^{2}) ds_{+} ds_{-}},
\end{equation}
where the sum of fractions over all decay channels may not be $100\%$ due to interference. Table~\ref{tab_fraction} summarizes the relative fractions for all solutions.
\begin{table}[htb]
  \caption{Time-dependent Dalitz plot fit result. The phases, $b_{i}$ and $d_{i}$, are given in radians.}
  \label{tab_dtcpv}
  \begin{tabular}
    {@{\hspace{0.5cm}}c@{\hspace{0.5cm}}| @{\hspace{0.5cm}}c@{\hspace{0.25cm}}  @{\hspace{0.25cm}}c@{\hspace{0.25cm}}  @{\hspace{0.25cm}}c@{\hspace{0.25cm}}  @{\hspace{0.25cm}}c@{\hspace{0.5cm}}}
    \hline \hline
    \begin{sideways}Parameter\end{sideways} & \begin{sideways}Solution 1\end{sideways} & \begin{sideways}Solution 2\end{sideways} & \begin{sideways}Solution 3\end{sideways} & \begin{sideways}Solution 4\end{sideways} \\
    \hline
    $c_{\Ksp}$                 & $0.106 \pm 0.058$ & $0.099 \pm 0.057$ & $0.081 \pm 0.058$ & $0.116 \pm 0.058$ \\
    $d_{\Ksp}$                 & $-0.006 \pm 0.200$ & $0.127 \pm 0.172$ & $-0.013 \pm 0.190$ & $0.227 \pm 0.167$ \\
    $a_{\Kstarp}$              & $71.373 \pm 5.147$ & $38.553 \pm 4.845$ & $42.430 \pm 5.800$ & $48.990 \pm 5.035$ \\
    $b_{\Kstarp}$              & $-2.507 \pm 0.144$ & $-2.983 \pm 0.153$ & $3.294 \pm 0.144$ & $-2.829 \pm 0.149$ \\
    $c_{\Kstarp}$              & $-0.003 \pm 0.030$ & $0.081 \pm 0.086$ & $0.126 \pm 0.096$ & $0.259 \pm 0.096$ \\
    $d_{\Kstarp}$              & $0.397 \pm 0.171$ & $0.636 \pm 0.156$ & $0.483 \pm 0.193$ & $0.874 \pm 0.157$ \\
    $a_{\rhoz}$                & $1.163 \pm 0.133$ & $1.394 \pm 0.183$ & $1.737 \pm 0.145$ & $1.200 \pm 0.169$ \\
    $b_{\rhoz}$                & $-2.528 \pm 0.306$ & $-3.934 \pm 0.394$ & $1.418 \pm 0.229$ & $-3.479 \pm 0.523$ \\
    $c_{\rhoz}$                & $-0.017 \pm 0.115$ & $0.083 \pm 0.119$ & $0.035 \pm 0.079$ & $0.057 \pm 0.129$ \\
    $d_{\rhoz}$                & $0.350 \pm 0.148$ & $0.398 \pm 0.131$ & $0.406 \pm 0.136$ & $0.457 \pm 0.141$ \\
    $a_{\fz}$                  & $31.840 \pm 2.697$ & $32.957 \pm 2.837$ & $46.640 \pm 3.527$ & $32.410 \pm 2.707$ \\
    $b_{\fz}$                  & $-1.309 \pm 0.258$ & $-1.973 \pm 0.284$ & $-2.783 \pm 0.213$ & $-1.850 \pm 0.292$ \\
    $c_{\fz}$                  & $0.032 \pm 0.083$ & $-0.000 \pm 0.084$ & $0.006 \pm 0.055$ & $-0.005 \pm 0.085$ \\
    $d_{\fz}$                  & $0.221 \pm 0.115$ & $0.258 \pm 0.121$ & $0.259 \pm 0.100$ & $0.248 \pm 0.117$ \\
    $a_{\ftwo}$                & $0.101 \pm 0.020$ & $0.121 \pm 0.021$ & $ 0.091 \pm 0.022$ & $0.116 \pm 0.021$ \\
    $b_{\ftwo}$                & $-0.303 \pm 0.235$ & $-0.880 \pm 0.217$ & $-0.710 \pm 0.311$ & $-0.829 \pm 0.240$ \\
    $c_{\rm Rest}$             & $-0.042 \pm 0.043$ & $-0.016 \pm 0.043$ & $-0.022 \pm 0.037$ & $-0.029 \pm 0.045$ \\
    $d_{\rm Rest}$             & $0.438 \pm 0.174$ & $0.6053 \pm 0.121$ & $0.380 \pm 0.116$ & $0.558 \pm 0.115$ \\
    $a_{\fX}$                  & $8.303 \pm 1.394$ & $13.748 \pm 1.595$ & $12.560 \pm 2.005$ & $13.080 \pm 1.577$ \\
    $b_{\fX}$                  & $0.382 \pm 0.288$ & $0.009 \pm 0.234$ & $0.578 \pm 0.278$ & $0.047 \pm 0.253$ \\
    $a_{(\Ks \pip)_{\rm NR}}$  & $109.661 \pm 9.777$ & $104.821 \pm 10.038$ & $114.600 \pm 9.228$ & $103.400 \pm 9.538$ \\
    $b_{(\Ks \pip)_{\rm NR}}$  & $-3.591 \pm 0.129$ & $-3.165 \pm 0.149$ & $3.163 \pm 0.155$ & $-3.396 \pm 0.151$ \\
    $a_{(\Ks \pim)_{\rm NR}}$  & $24.661 \pm 20.171$ & $35.387 \pm 9.859$ & $57.170 \pm 12.660$ & $42.490 \pm 11.270$ \\
    $b_{(\Ks \pim)_{\rm NR}}$  & $-1.984 \pm 0.430$ & $-1.531 \pm 0.331$ & $5.251 \pm 0.209$ & $-1.779 \pm 0.245$ \\
    $a_{(\pip \pim)_{\rm NR}}$ & $42.197 \pm 6.620$ & $34.769 \pm 9.004$ & $53.560 \pm 8.202$ & $40.960 \pm 7.972$ \\
    $b_{(\pip \pim)_{\rm NR}}$ & $-0.230 \pm 0.184$ & $-0.515 \pm 0.197$ & $1.118 \pm 0.216$ & $-0.556 \pm 0.242$ \\
    $\alpha$                   & $0.147 \pm 0.032$ & $0.134 \pm 0.020$ & $0.089 \pm 0.025$ & $0.127 \pm 0.020$ \\ \hline
    $-2 \log {\cal L}$         & $18472.5$ & $18465.0$ & $18458.5$ & $18465.9$ \\
    \hline \hline
  \end{tabular}
\end{table}
\begin{table}
  \caption{Summary of relative fractions where only the statistical errors are given.}
  \label{tab_fraction}
  \begin{tabular}
    {@{\hspace{0.5cm}}c@{\hspace{0.5cm}}||@{\hspace{0.5cm}}c@{\hspace{0.5cm}}  @{\hspace{0.5cm}}c@{\hspace{0.5cm}}  @{\hspace{0.5cm}}c@{\hspace{0.5cm}}  @{\hspace{0.5cm}}c@{\hspace{0.5cm}}}
    \hline \hline
    \begin{sideways}Decay channel\end{sideways} & \begin{sideways}Sol. 1 Fraction (\%)\end{sideways} & \begin{sideways}Sol. 2 Fraction (\%)\end{sideways} & \begin{sideways}Sol. 3 Fraction (\%)\end{sideways} & \begin{sideways}Sol. 4 Fraction (\%)\end{sideways}\\
    \hline
    \Ksp \pim& $9.3 \pm 0.8$ & $9.0 \pm 1.3$ & $8.7 \pm 1.2$ & $9.1 \pm 1.2$\\
    \Kstarp \pim & $61.7 \pm 10.4$ & $17.4 \pm 5.0$ & $20.5 \pm 6.4$ & $28.5 \pm 7.4$\\
    \rhoz \Ks& $6.1 \pm 1.5$ & $8.5 \pm 2.6$ & $12.8 \pm 2.8$ & $6.4 \pm 2.0$\\
    \fz \Ks& $14.3 \pm 2.7$ & $14.9 \pm 3.3$ & $28.9 \pm 5.9$ & $14.6 \pm 3.1$\\
    \ftwo \Ks& $2.6 \pm 0.9$ & $3.2 \pm 1.2$ & $1.8 \pm 0.9$ & $3.1 \pm 1.1$\\
    \fX \Ks& $2.3 \pm 0.8$& $6.0 \pm 1.6$ & $4.9 \pm 1.7$ & $5.5 \pm 1.5$\\
    $(\Ks \pip)_{\rm NR} \pim$ & $57.2 \pm 11.4$ & $55.9 \pm 13.3$ & $96.7 \pm 20.5$ & $58.2 \pm 13.2$\\
    $(\Ks \pim)_{\rm NR} \pip$ & $2.9 \pm 4.7$ & $6.4 \pm 3.7$ & $24.1 \pm 11.2$ & $9.8 \pm 5.4$\\
    $(\pip \pim)_{\rm NR} \Ks$ & $10.7 \pm 3.5$ & $7.7 \pm 4.1$ & $24.9 \pm 8.4$ & $11.3 \pm 4.7$\\
    \hline
    Total & $167.1 \pm 0.2$ & $129.0 \pm 0.2$ & $223.2 \pm 0.3$ & $146.5 \pm 0.17$\\
    \hline \hline
  \end{tabular}
\end{table}

Solutions 1 and 2 were found in the previous Belle analysis~\cite{kspipi_Belle1} and correspond to the two solutions found in the higher statistics Belle Dalitz plot analysis of the isospin partner mode, $\Bp \rightarrow \Kp \pip \pim$~\cite{kpipi_Belle}. The signal models used in this analysis and that of the isospin partner are almost identical. These two solutions are due to an interplay between the two broad $S$-wave amplitudes, $\Kstarp \pim$ and the non-resonant component, and are characterized by their different relative fractions. However, they give almost indistinguishable Dalitz plot distributions; studies with high statistics pseudo-experiments confirm the existence of these two solutions. Solution 3 has a high $\fz \Ks$ fraction compared with Solution 2 that may indicate a similar type of interference between \fz\ and the non-resonant component. Unlike the previous case, Solution 3 cannot be found in a high statistics toy MC sample generated with Solution 2, neither can Solution 2 be reproduced from Solution 3 indicating that one of these solutions cannot survive with higher experimental statistics. Considering that a solution with a high $\fz \Ks$ fraction was not found in the previous analysis or the higher statistics $\Bp \rightarrow \Kp \pip \pim$ analysis~\cite{kpipi_Belle}, we therefore conclude that Solution 3 is due to statistical fluctuations in the small \Kspipi\ sample and is unlikely to be a physical solution. Solution 4 appears to be statistically consistent with Solution 2 for all parameters, while Solution 4 cannot be reproduced from a high statistics toy MC sample generated with Solution 2 and vice versa. This indicates that these two solutions are likely due to a statistical fluctuation of a single solution and we select Solution 2 as the nominal solution of the two because of its slightly better likelihood.

Consequently, Solutions 1 and 2 are treated as the solutions of this analysis. The high $\Kstarp \pim$ fraction of Solution 1 is in agreement with some phenomenological estimates~\cite{Kstarpi} and may also be qualitatively favored by the total $K-\pi$ $S$-wave phase shift as a function of $m(K\pi)$ when compared with that measured by the LASS collaboration~\cite{LASS}.  As the likelihood difference is not found to be significant using an ensemble of pseudo-experiments corresponding to the luminosity of the data sample, we retain both solutions.

\begin{figure}[htb]
  \centering
  \includegraphics[height=160pt,width=!]{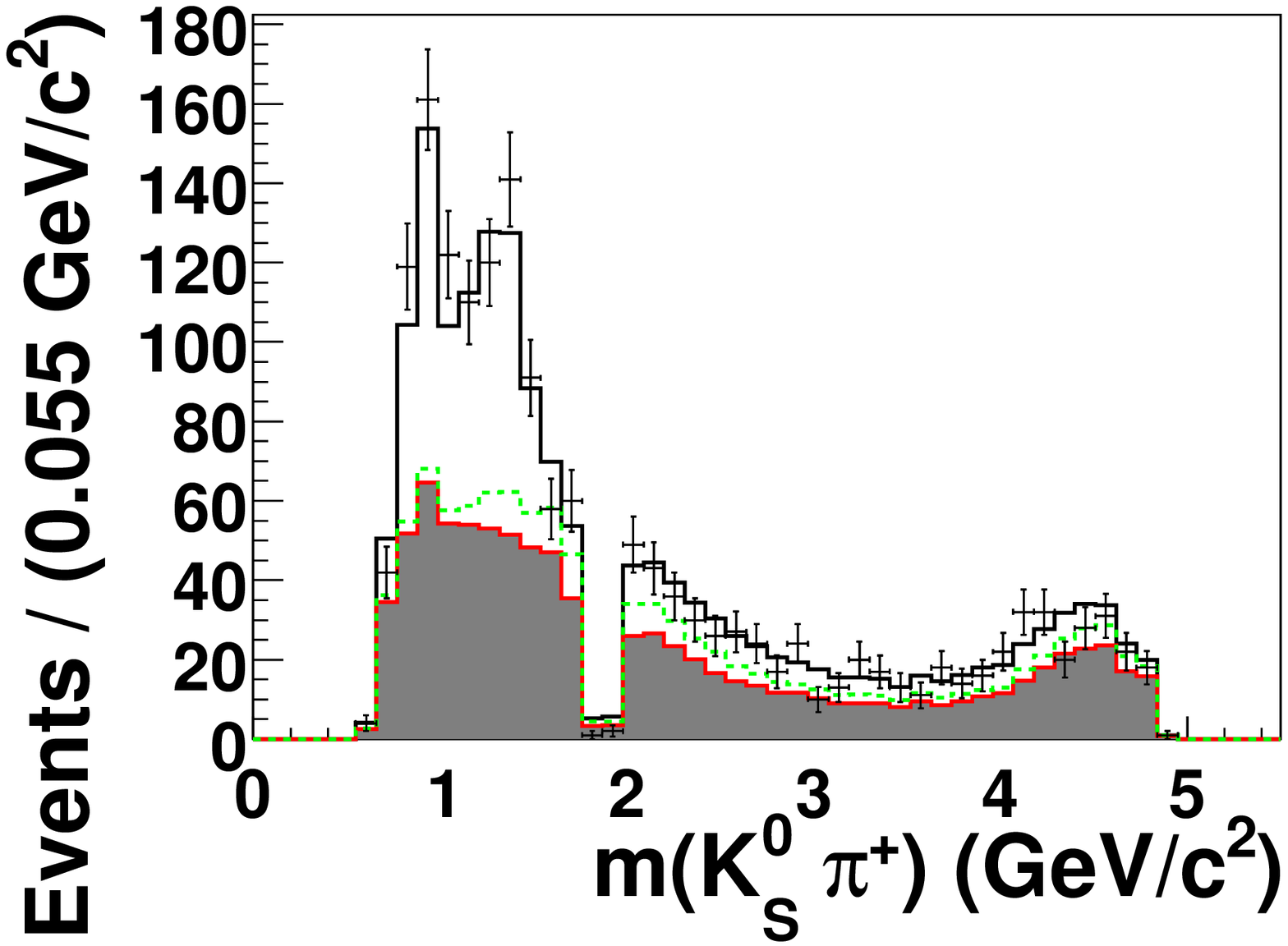}
  \includegraphics[height=160pt,width=!]{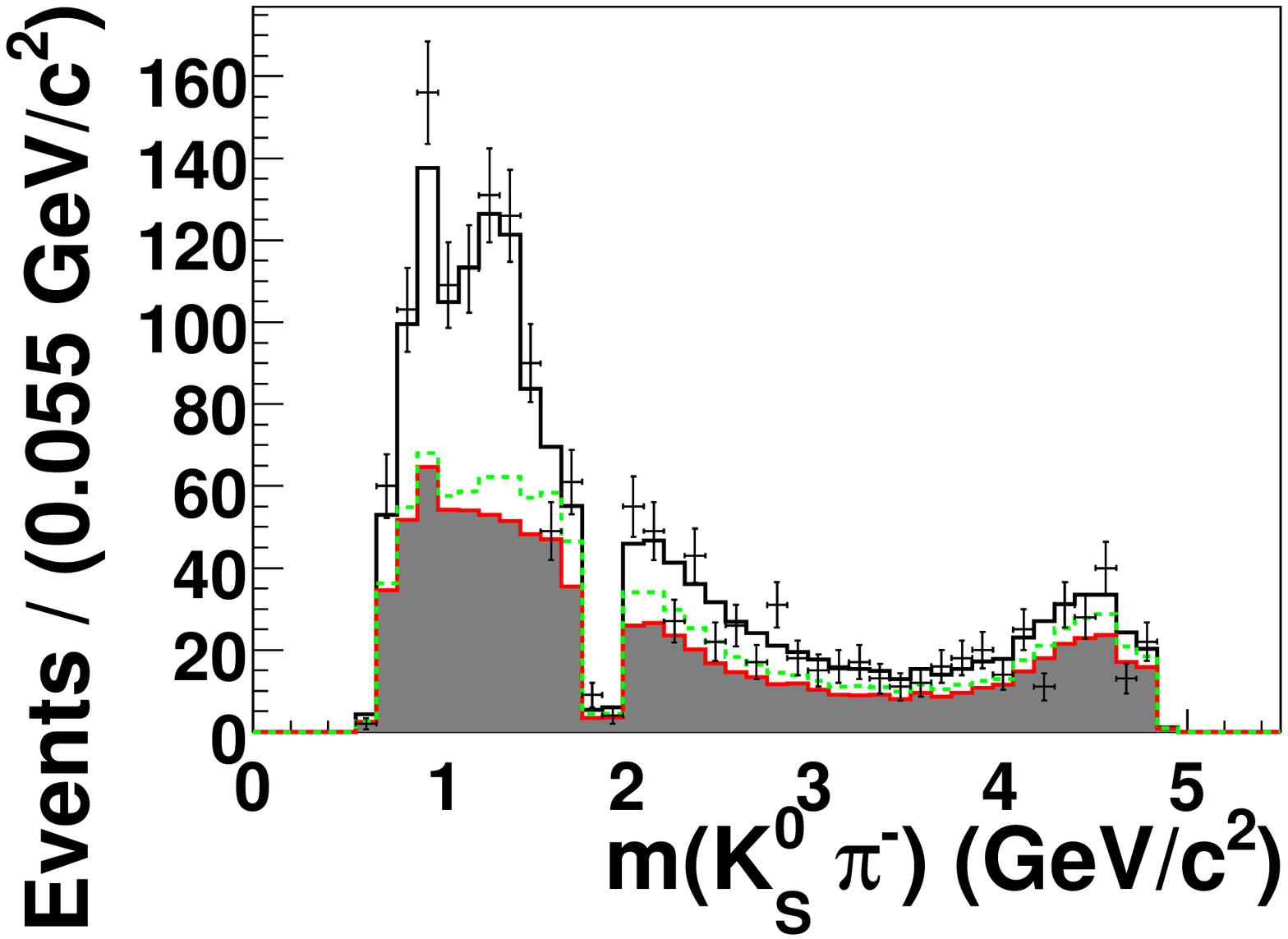}

  \includegraphics[height=160pt,width=!]{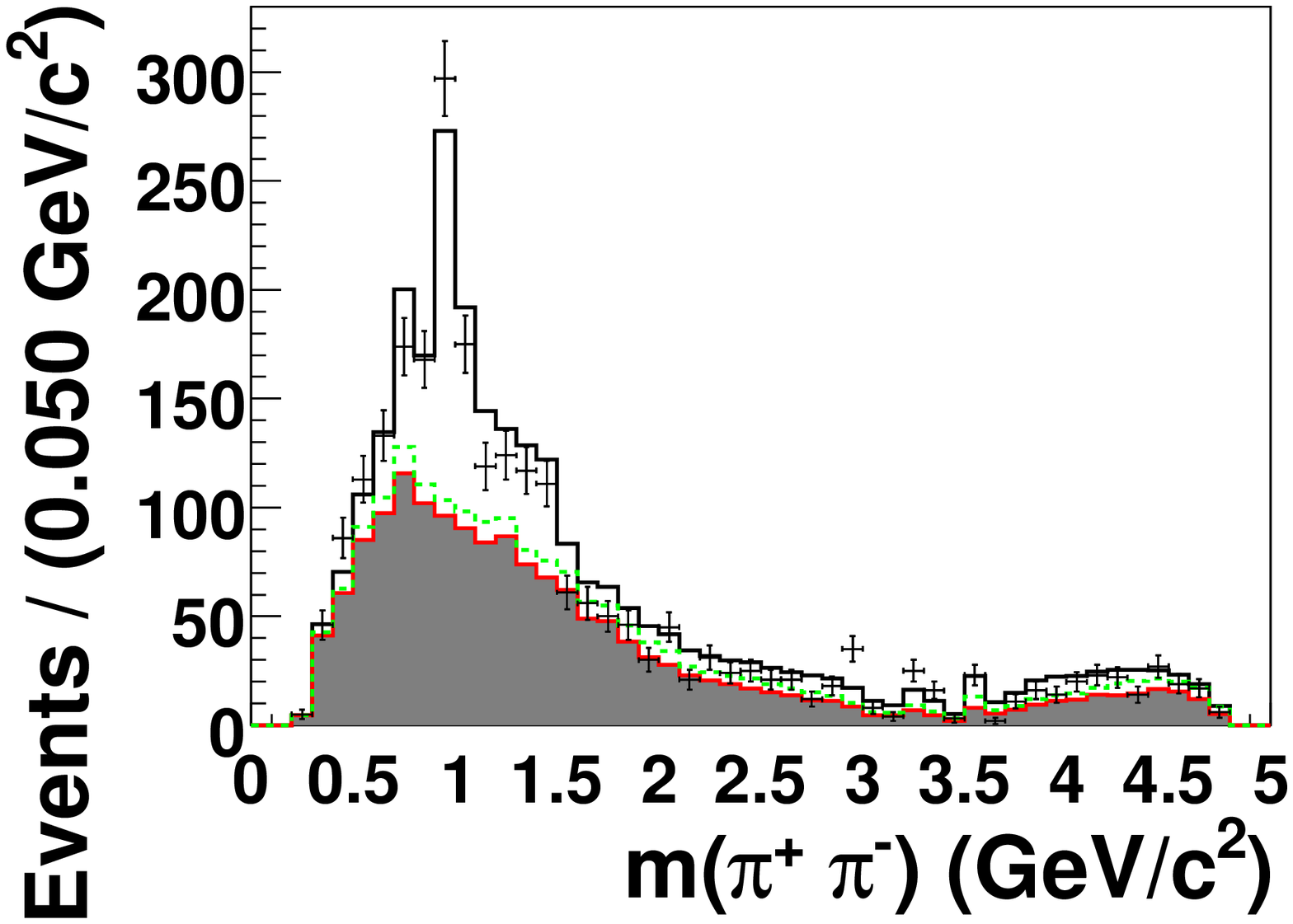}
  \caption{Signal enhanced Dalitz plot fit for \Kspipi. When plotting a two-particle mass projection, we require the invariant mass of the other two two-particle combinations to be greater than $1.5 \; {\rm GeV}/c^{2}$. The solid curve shows the fit result while the shaded and dashed curves show the \qqbar\ and total background components, respectively.}
  \label{fig_dp}
\end{figure}

To assess how well the fit to the Dalitz plot represents data as shown in Fig.~\ref{fig_dp}, the time and flavor-integrated Dalitz plot is divided into variable size bins so that each bin contains at least 25 events. A goodness-of-fit statistic for the multinomial distribution is then calculated as the sum over $N$ bins,
\begin{equation}
  \chi^{2} = -2 \sum^{N}_{i=1}n_{i} \log\frac{p_{i}}{n_{i}},
\end{equation}
where $n_{i}$ is the number of events observed in the $i$-th bin and $p_{i}$ is the number of events in the $i$-th bin as given by the fit~\cite{chisq}. The distribution of this statistic is bounded by a $\chi^{2}$ distribution with $N-1$ degrees of freedom and one with $N-k-1$ degrees of freedom where $k=19$ is the number of Dalitz plot fit parameters. The $\chi^{2}/N$ value for the best fit is $201.9/(189-19-1)$.

We measure the time-dependent $CP$ parameters of \Kspipi,

Solution 1: $-2 \log {\cal L} = 18472.5$
\begin{eqnarray}
  \Acp(\rhoz \Ks) &=& +0.03^{+0.23}_{-0.24} \pm 0.11 \pm 0.10, \nonumber \\
  \phione(\rhoz \Ks) &=& (+20.0^{+8.6}_{-8.5} \pm 3.2 \pm 3.5)^{\circ}, \nonumber \\
  \Scp(\rhoz \Ks) &=& +0.64^{+0.19}_{-0.25} \pm 0.09 \pm 0.10, \nonumber \\
  \Acp(\fz \Ks) &=& -0.06 \pm 0.17 \pm 0.07 \pm 0.09, \nonumber \\
  \phione(\fz \Ks) &=& (+12.7^{+6.9}_{-6.5} \pm 2.8 \pm 3.3)^{\circ}, \nonumber \\
  \Scp(\fz \Ks) &=& -0.43^{+0.22}_{-0.20} \pm 0.09 \pm 0.11, \nonumber \\
  \Acp(\Ksp \pim) &=& -0.21 \pm 0.11 \pm 0.05 \pm 0.05, \nonumber \\
  \Delta \phi(\Ksp \pim) &=& (-0.7^{+23.5}_{-22.8} \pm 11.0 \pm 17.6),^{\circ}
\end{eqnarray}

\begin{figure}[htb]
  \centering
  \includegraphics[height=280pt,width=!]{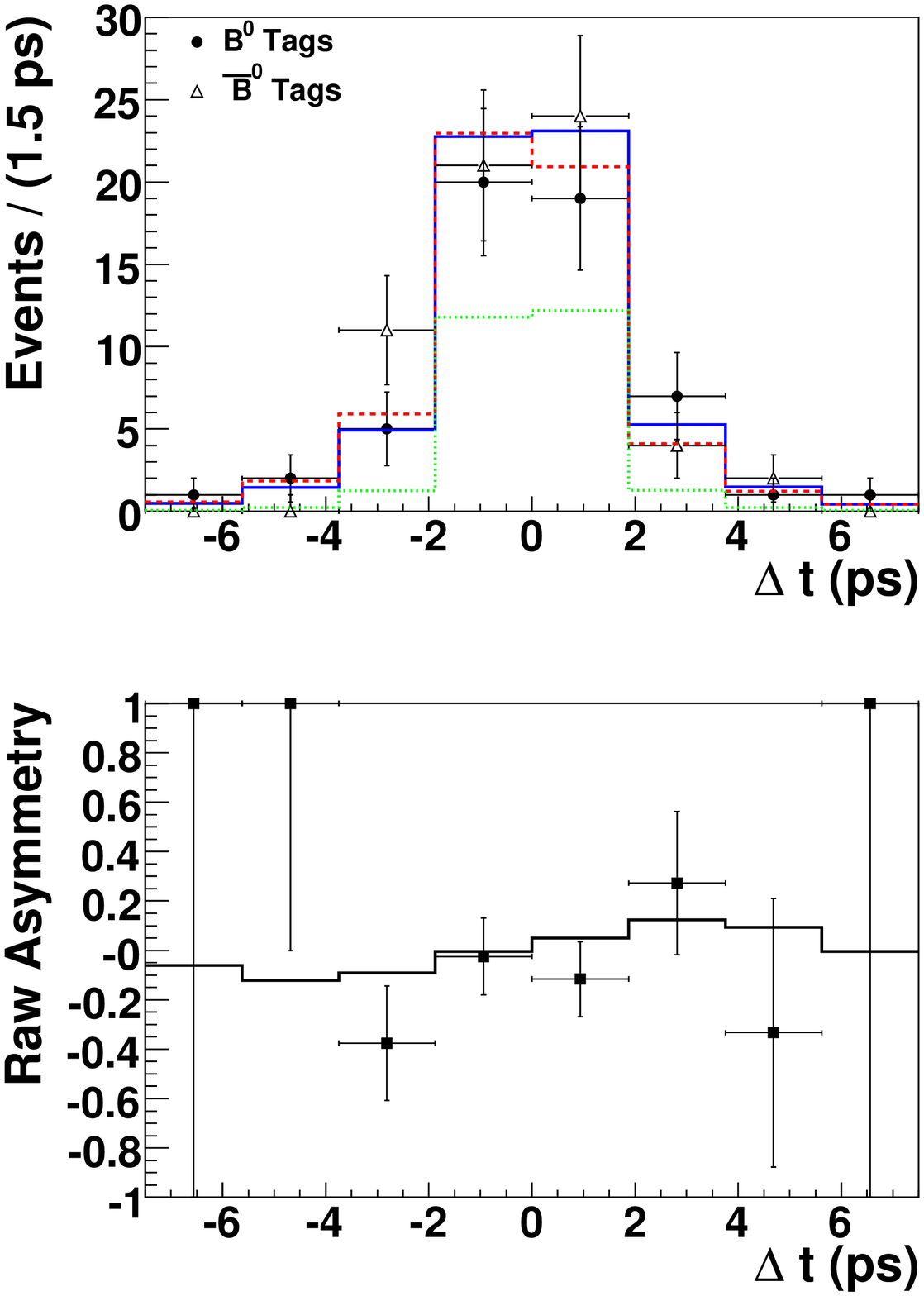}
  \includegraphics[height=280pt,width=!]{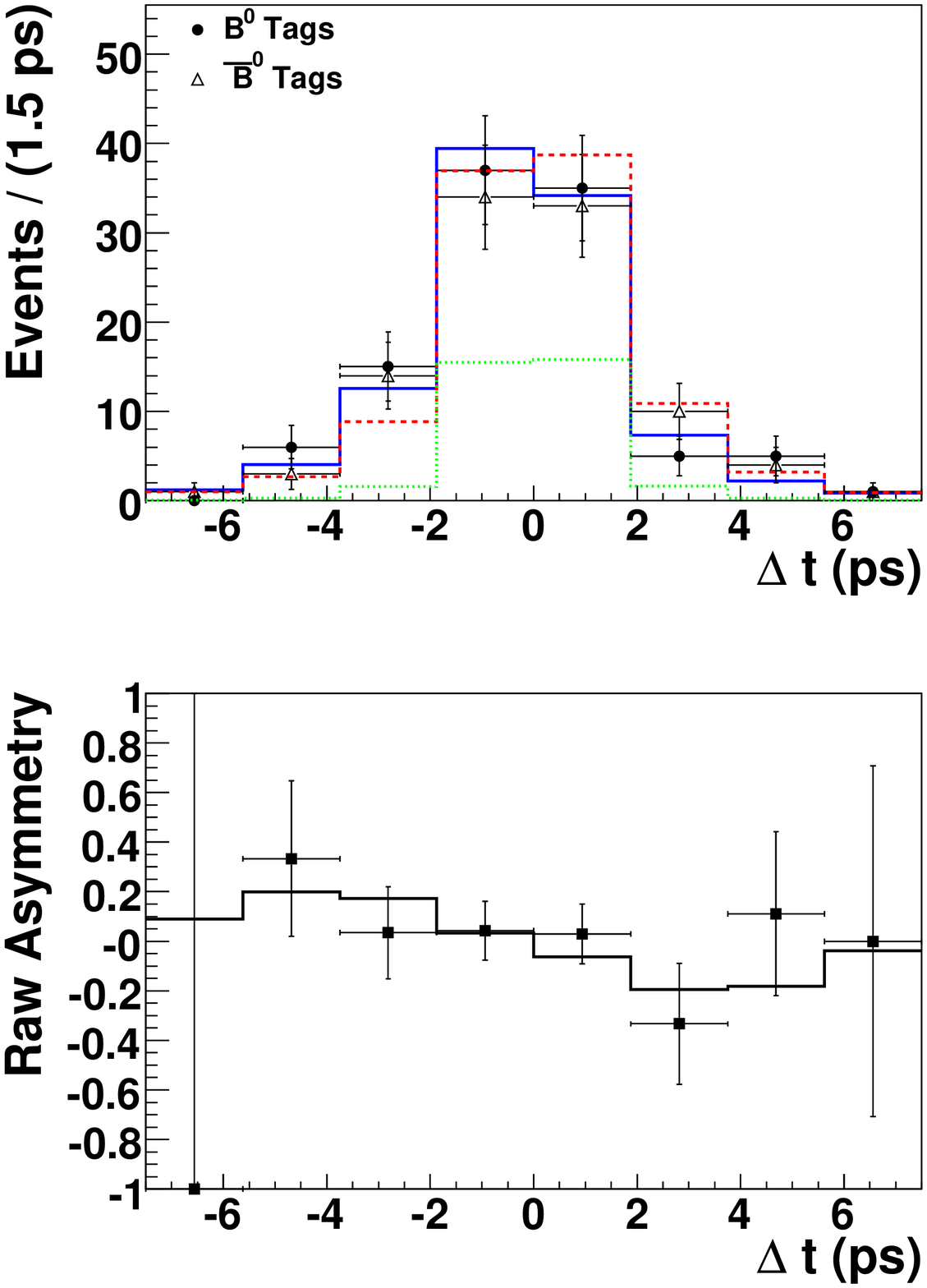}
  \put(-261,250){(a)}
  \put(-43,250){(b)}
  \caption{Time-dependent Dalitz plot fit results for \Kspipi\ in (a), the \rhoz\ region and (b), the \fz\ region using Solution 1. The top plots show the \Dt\ distribution for \Bz\ and \Bzb\ tags indicated by the solid and dashed curves, respectively. These plots contain only good tags, $0.5 < r \leq 1.0 $ and the dotted curve represents the background contribution. The bottom plots show the \BzBzb\ raw asymmetry of the top plots, $(N_{\Bz} - N_{\Bzb})/(N_{\Bz} + N_{\Bzb})$, where $N_{\Bz} \, (N_{\Bzb})$ is the number of \Bz\ (\Bzb) tags in \Dt.}
  \label{fig_dpdt}
\end{figure}

Solution 2: $-2 \log {\cal L} = 18465.0$
\begin{eqnarray}
  \Acp(\rhoz \Ks) &=& -0.16 \pm 0.24 \pm 0.12 \pm 0.10, \nonumber \\
  \phione(\rhoz \Ks) &=& (+22.8 \pm 7.5 \pm 3.3 \pm 3.5)^{\circ}, \nonumber \\
  \Scp(\rhoz \Ks) &=& +0.71^{+0.15}_{-0.20} \pm 0.08 \pm 0.09, \nonumber \\
  \Acp(\fz \Ks) &=& +0.00 \pm 0.17 \pm 0.06 \pm 0.09, \nonumber \\
  \phione(\fz \Ks) &=& (+14.8^{+7.3}_{-6.7} \pm 2.7 \pm 3.3)^{\circ}, \nonumber \\
  \Scp(\fz \Ks) &=& -0.49^{+0.22}_{-0.20} \pm 0.08 \pm 0.10, \nonumber \\
  \Acp(\Ksp \pim) &=& -0.20 \pm 0.11 \pm 0.05 \pm 0.05, \nonumber \\
  \Delta \phi(\Ksp \pim) &=& (+14.6^{+19.4}_{-20.3} \pm 11.0 \pm 17.6),^{\circ}
\end{eqnarray}
where the first error is statistical, the second is systematic and the third is the Dalitz plot signal model uncertainty. The \Dt\ and raw asymmetry fit projections for \rhozks\ and \fzks\ are shown in Fig.~\ref{fig_dpdt}, and the statistical correlation coefficients between the $CP$ parameters of both solutions are given in Tables~\ref{tab_dtcpv_corr1} and \ref{tab_dtcpv_corr2}. The full correlation matrices are given in Tables~\ref{tab_corr11}-\ref{tab_corr23}.
\begin{table}[htb]
  \caption{Correlation between $CP$ parameters for Solution 1.}
  \label{tab_dtcpv_corr1}
  \begin{tabular}
    {@{\hspace{0.5cm}}c@{\hspace{0.5cm}}| @{\hspace{0.5cm}}c@{\hspace{0.5cm}}  @{\hspace{0.5cm}}c@{\hspace{0.5cm}}  @{\hspace{0.5cm}}c@{\hspace{0.5cm}}  @{\hspace{0.5cm}}c@{\hspace{0.5cm}}  @{\hspace{0.5cm}}c@{\hspace{0.5cm}}  @{\hspace{0.5cm}}c@{\hspace{0.5cm}}}
    \hline \hline
    & \begin{sideways}$\Acp(\rhoz \Ks)$\end{sideways} & \begin{sideways}$\phione(\rhoz \Ks)$\end{sideways} & \begin{sideways}$\Acp(\fz \Ks)$\end{sideways} & \begin{sideways}$\phione(\fz \Ks)$\end{sideways} & \begin{sideways}$\Acp(\Ksp \pim)$\end{sideways} & \begin{sideways}$\Delta \phi(\Ksp \pim)$\end{sideways}\\
    \hline
    $\Acp(\rhoz \Ks)$  & $+1.00$\\
    $\phione(\rhoz \Ks)$ & $-0.01$ & $+1.00$\\
    $\Acp(\fz \Ks)$    & $-0.29$ & $-0.01$ & $+1.00$\\
    $\phione(\fz \Ks)$   & $-0.06$ & $+0.37$ & $+0.03$ & $+1.00$\\
    $\Acp(\Ksp \pim)$   & $+0.07$ & $+0.01$ & $+0.00$ & $-0.07$ & $+1.00$\\
    $\Delta \phi(\Ksp \pim)$ & $-0.09$ & $+0.15$ & $+0.03$ & $+0.19$ & $-0.06$ & $+1.00$\\
    \hline \hline
  \end{tabular}
\end{table}
\begin{table}[htb]
  \caption{Correlation between $CP$ parameters for Solution 2.}
  \label{tab_dtcpv_corr2}
  \begin{tabular}
    {@{\hspace{0.5cm}}c@{\hspace{0.5cm}}| @{\hspace{0.5cm}}c@{\hspace{0.5cm}}  @{\hspace{0.5cm}}c@{\hspace{0.5cm}}  @{\hspace{0.5cm}}c@{\hspace{0.5cm}}  @{\hspace{0.5cm}}c@{\hspace{0.5cm}}  @{\hspace{0.5cm}}c@{\hspace{0.5cm}}  @{\hspace{0.5cm}}c@{\hspace{0.5cm}}}
    \hline \hline
    & \begin{sideways}$\Acp(\rhoz \Ks)$\end{sideways} & \begin{sideways}$\phione(\rhoz \Ks)$\end{sideways} & \begin{sideways}$\Acp(\fz \Ks)$\end{sideways} & \begin{sideways}$\phione(\fz \Ks)$\end{sideways} & \begin{sideways}$\Acp(\Ksp \pim)$\end{sideways} & \begin{sideways}$\Delta \phi(\Ksp \pim)$\end{sideways}\\
    \hline
    $\Acp(\rhoz \Ks)$  & $+1.00$\\
    $\phione(\rhoz \Ks)$ & $-0.12$ & $+1.00$\\
    $\Acp(\fz \Ks)$    & $-0.40$ & $-0.05$ & $+1.00$\\
    $\phione(\fz \Ks)$   & $-0.13$ & $+0.43$ & $+0.14$ & $+1.00$\\
    $\Acp(\Ksp \pim)$   & $+0.02$ & $+0.00$ & $+0.01$ & $-0.05$ & $+1.00$\\
    $\Delta \phi(\Ksp \pim)$ & $+0.16$ & $+0.25$ & $-0.09$ & $+0.22$ & $+0.05$ & $+1.00$\\
    \hline \hline
  \end{tabular}
\end{table}

\begin{figure}[htb]
  \centering
  \includegraphics[height=150pt,width=!]{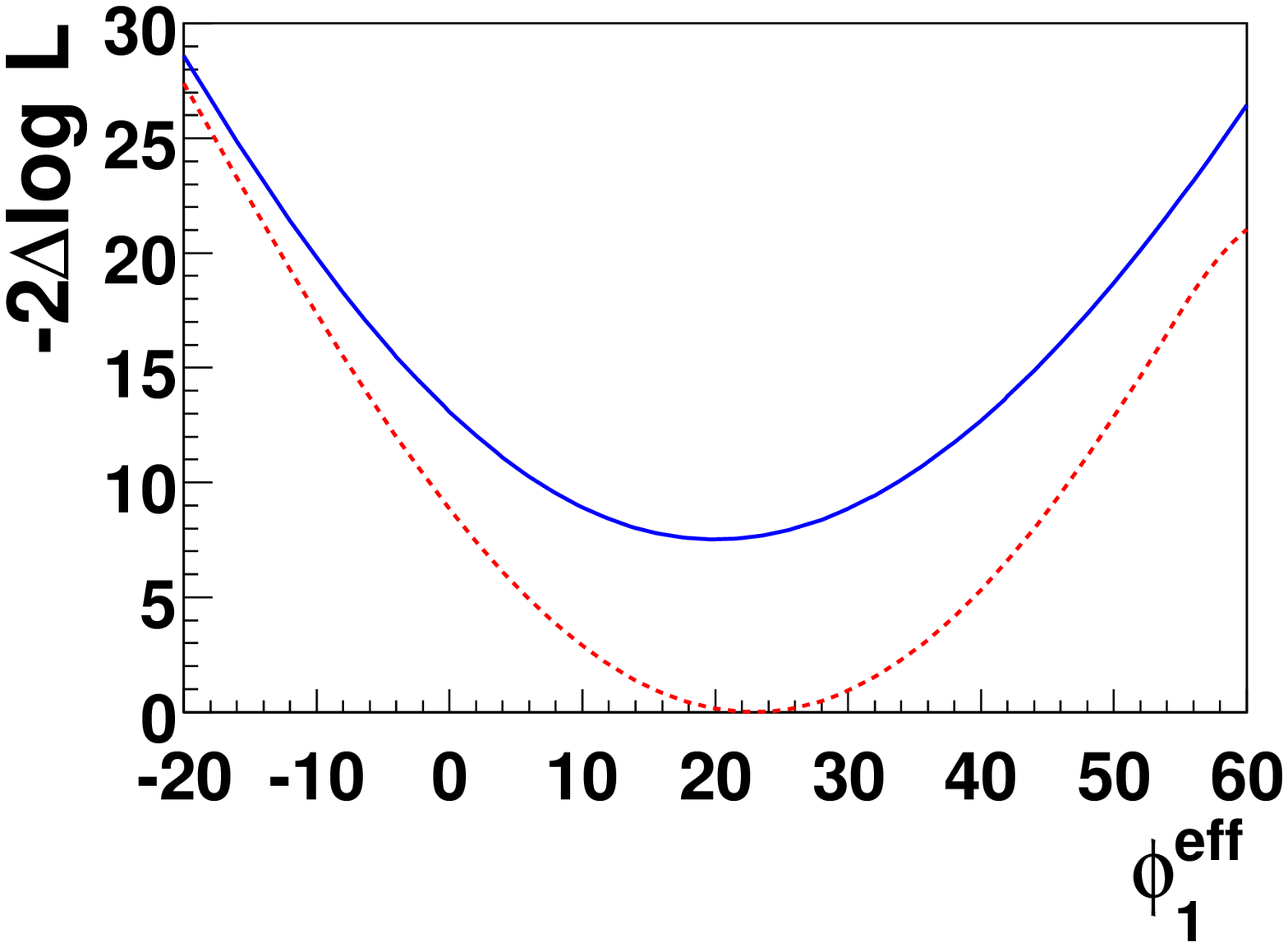}
  \includegraphics[height=150pt,width=!]{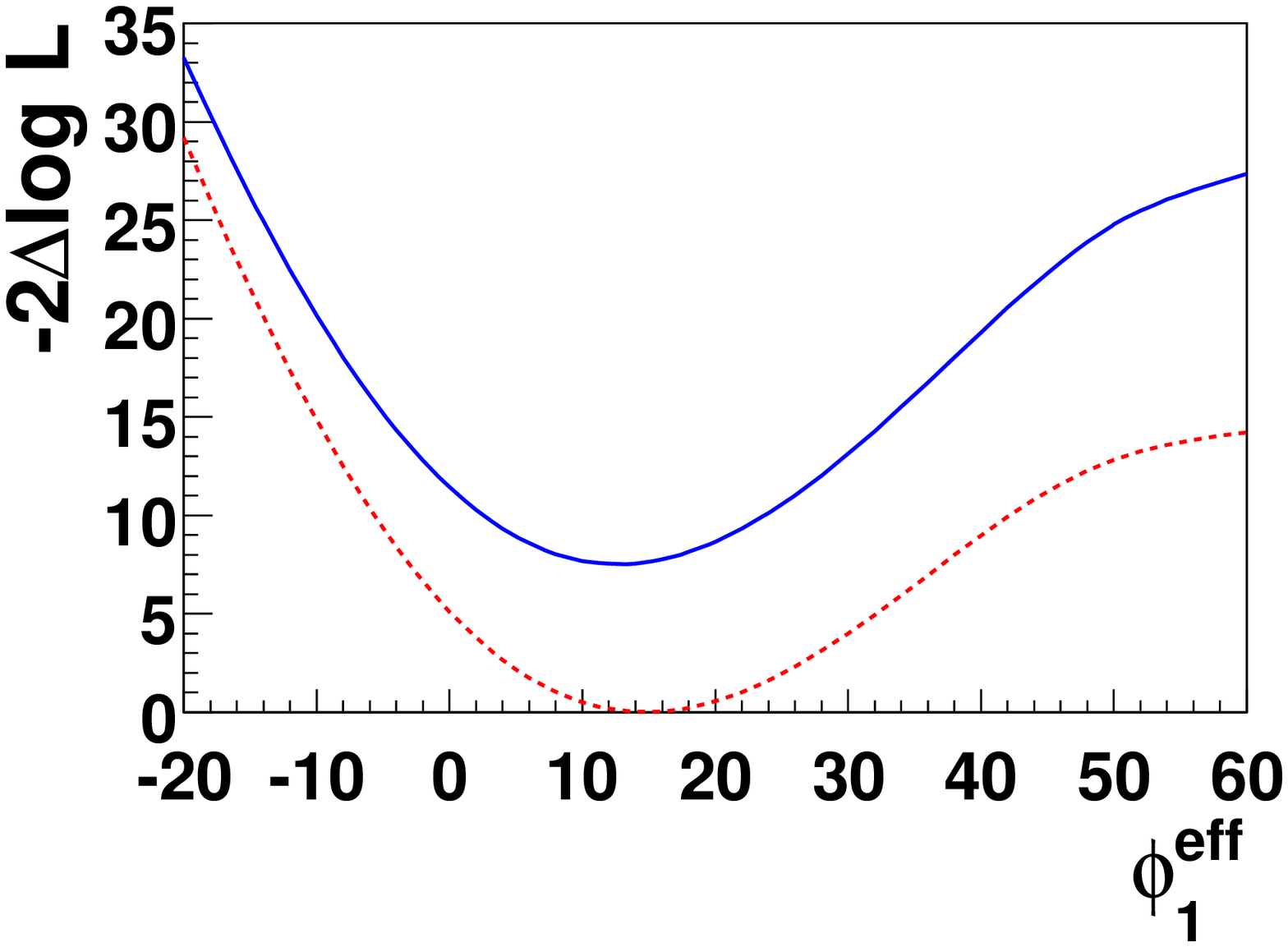}
  \put(-250,122){(a)}
  \put(-40,122){(b)}
  \caption{Statistical likelihood scan of \phione\ for \rhozks\ (a), and \fzks\ (b) where the solid (dashed) curve represents Solution 1 (2).}
  \label{fig_scan1}
\end{figure}

\begin{figure}[htb]
  \centering
  \includegraphics[height=150pt,width=!]{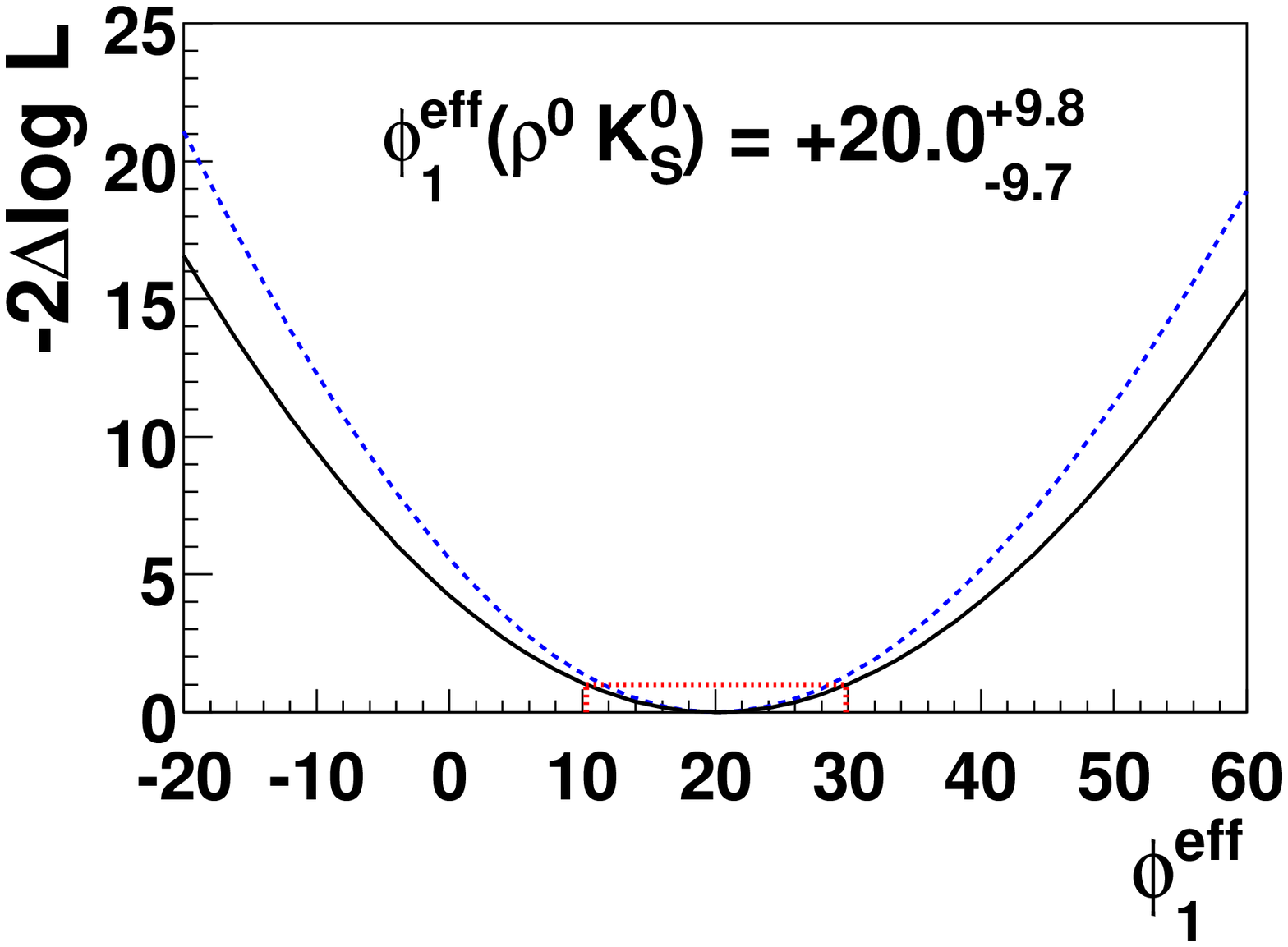}
  \includegraphics[height=150pt,width=!]{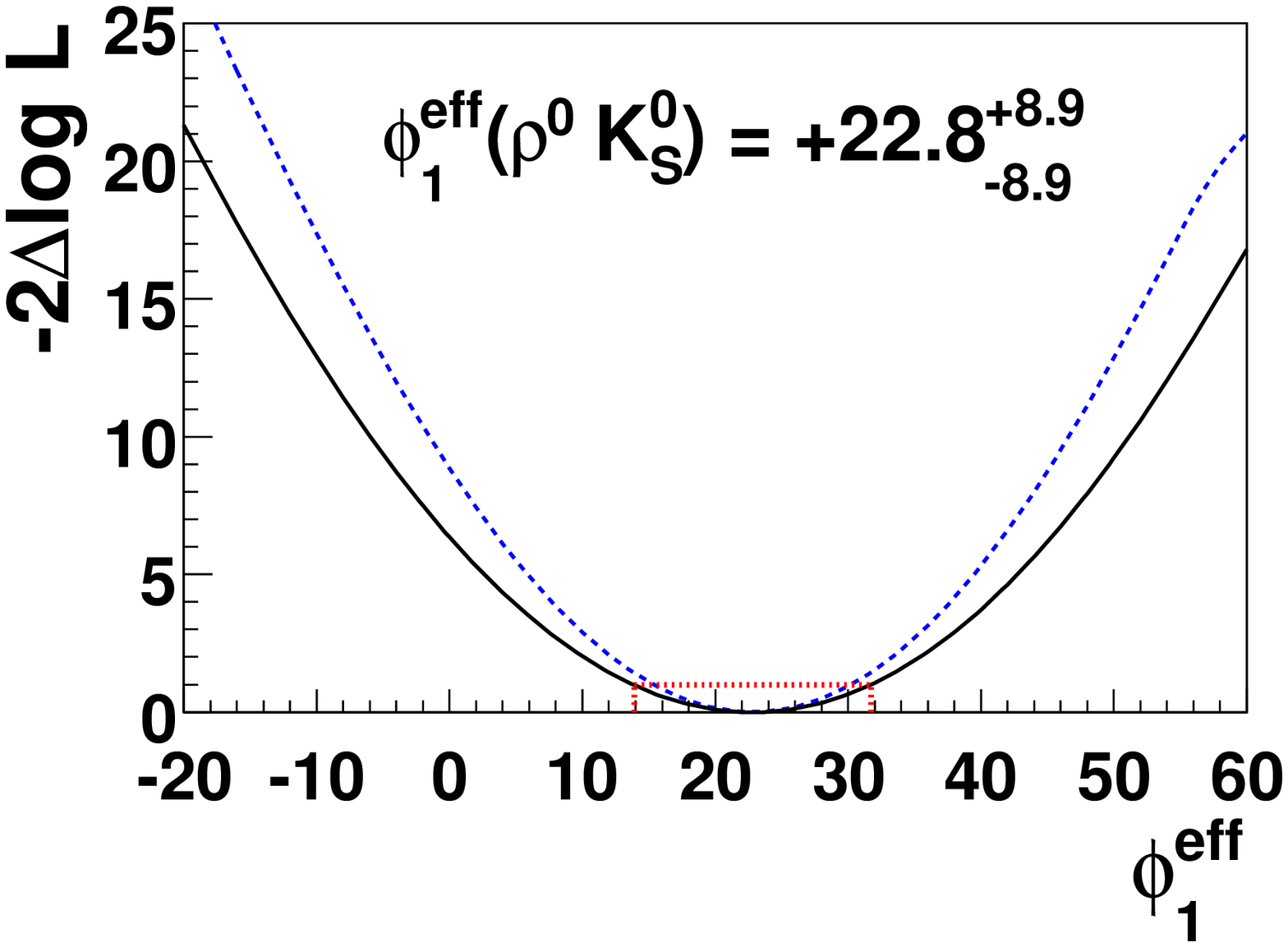}

  \includegraphics[height=150pt,width=!]{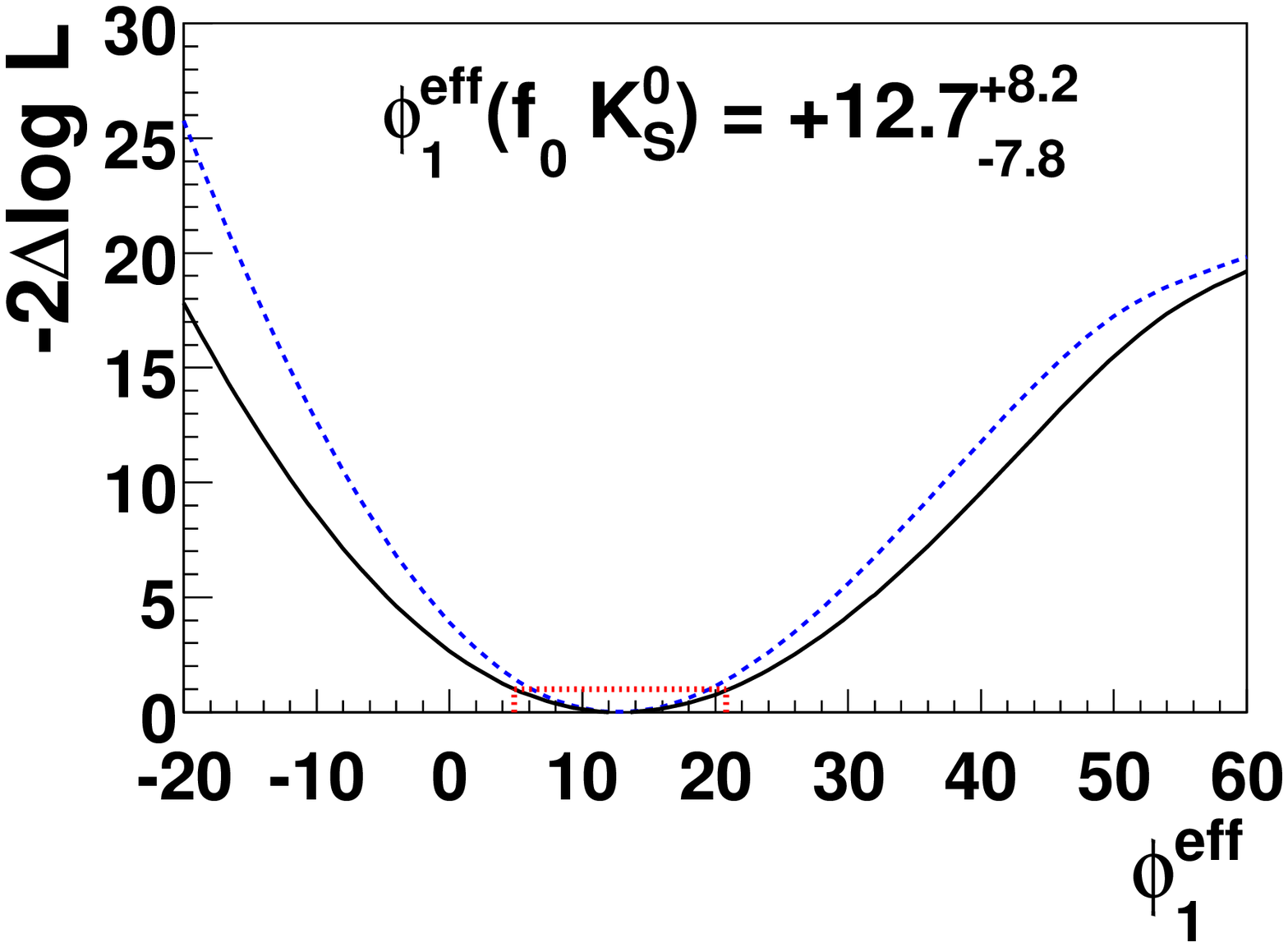}
  \includegraphics[height=150pt,width=!]{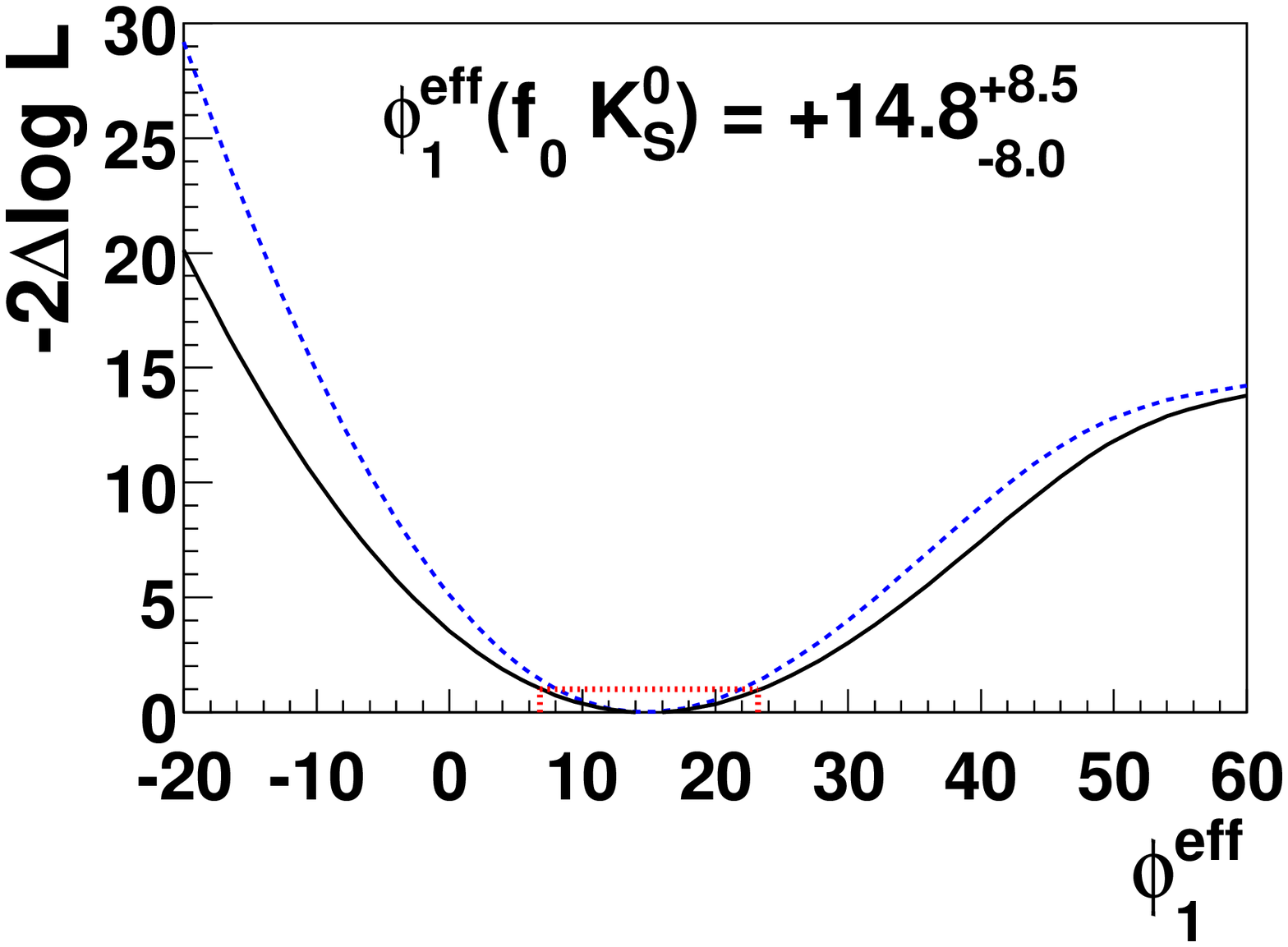}
  \caption{Likelihood scans of \phione\ for \rhozks\ (top) and \fzks\ (bottom) for Solution 1 (left) and Solution 2 (right). The solid (dashed) curve contains the total (statistical) error and the dotted box indicates the parameter range corresponding to $\pm 1 \sigma$.}
  \label{fig_scan2}
\end{figure}

\begin{figure}[htb]
  \centering
  \includegraphics[height=150pt,width=!]{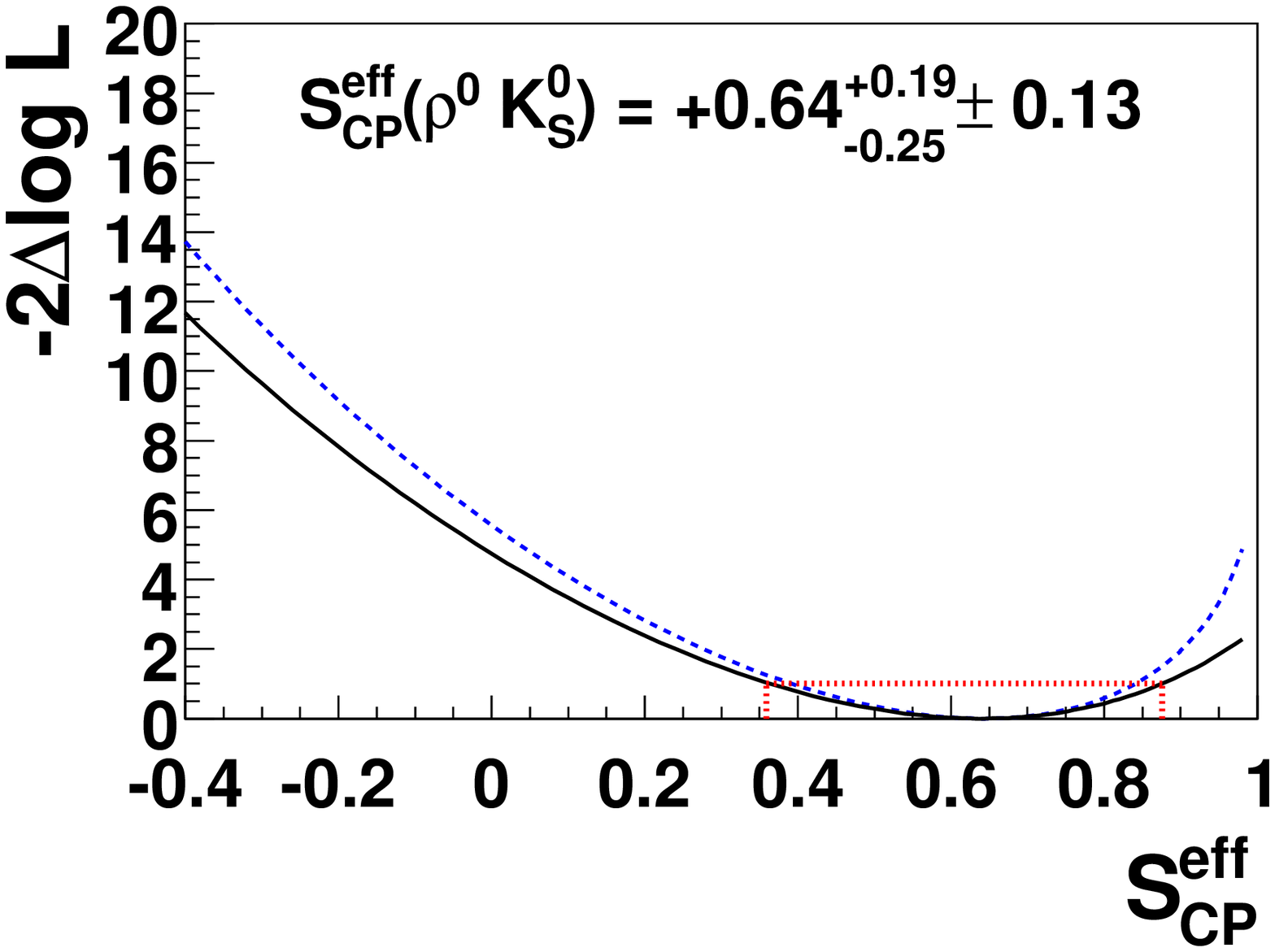}
  \includegraphics[height=150pt,width=!]{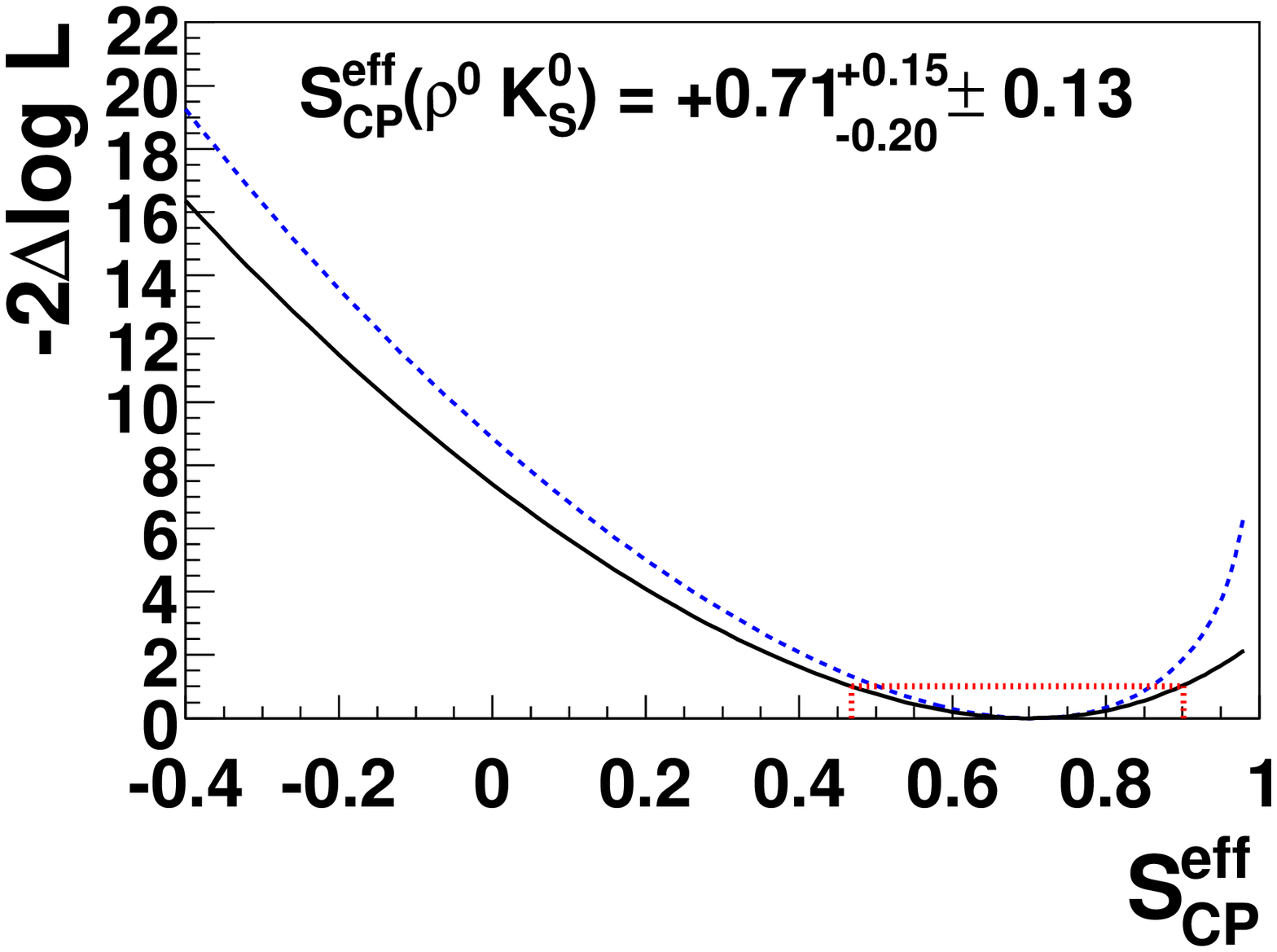}

  \includegraphics[height=150pt,width=!]{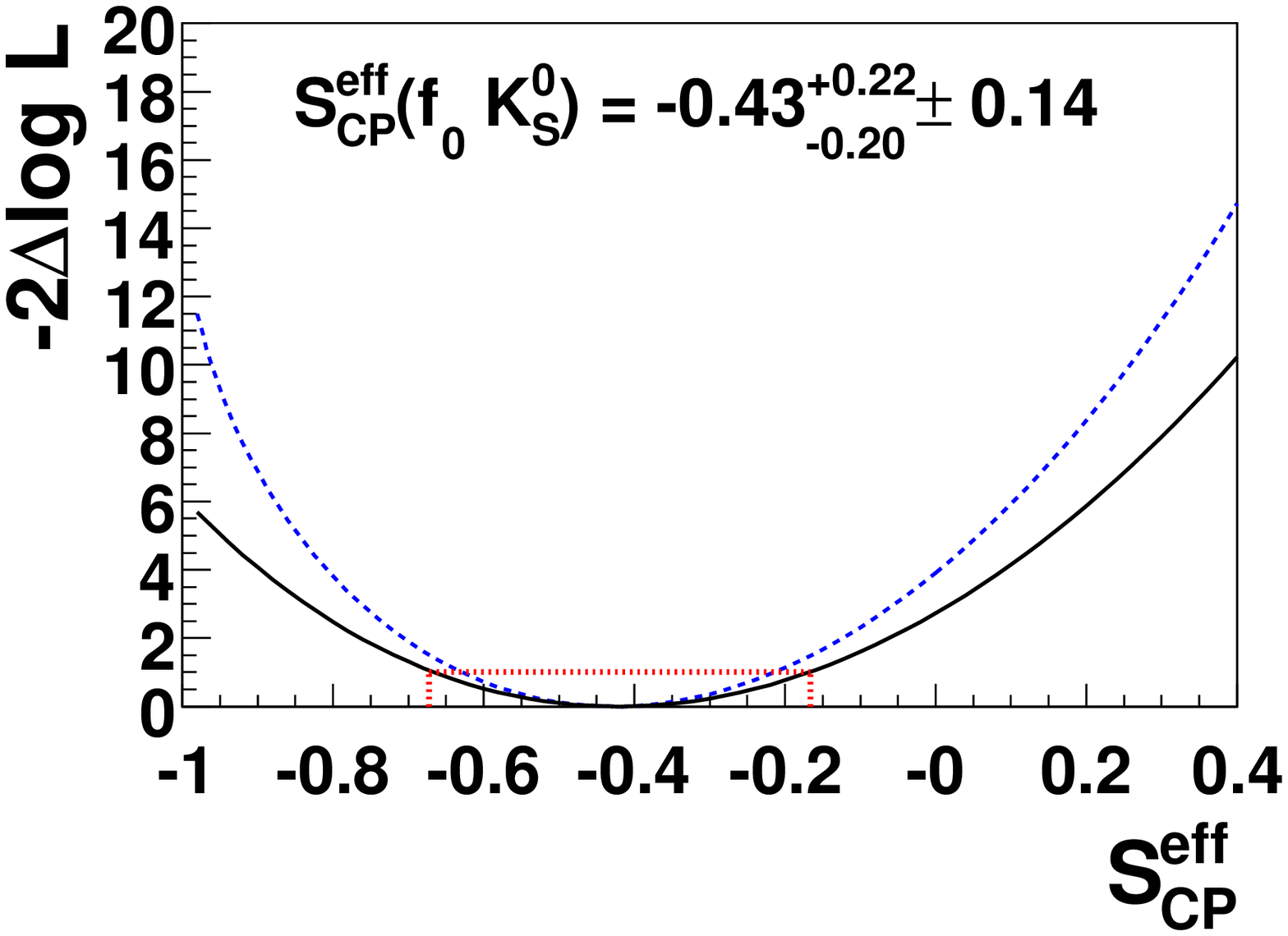}
  \includegraphics[height=150pt,width=!]{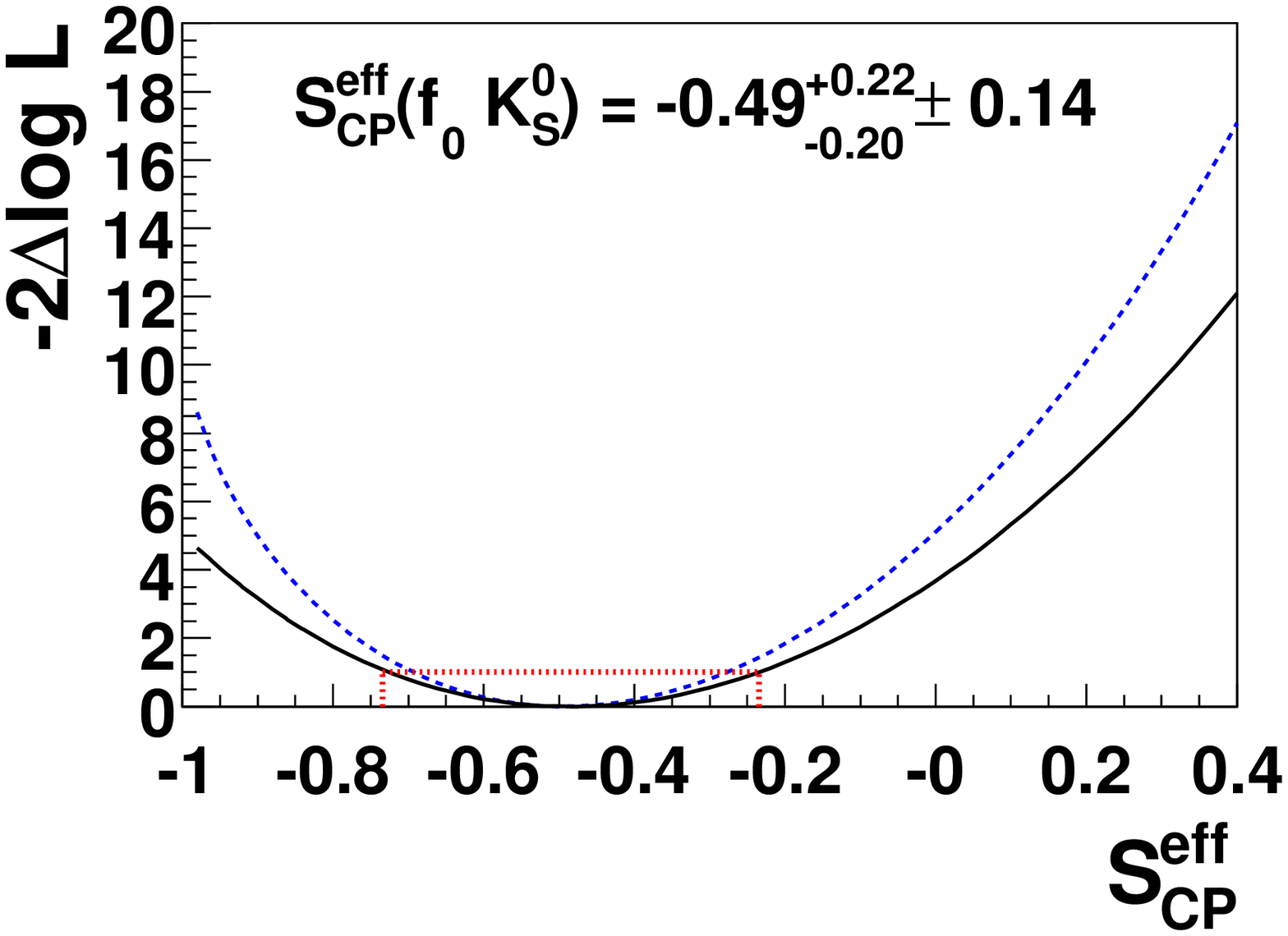}
  \caption{Likelihood scans of \Scp\ for \rhozks\ (top) and \fzks\ (bottom) for Solution 1 (left) and Solution 2 (right). The solid (dashed) curve contains the total (statistical) error and the dotted box indicates the parameter range corresponding to $\pm 1 \sigma$. The statistical error of \Scp\ is determined from these scans.}
  \label{fig_scan3}
\end{figure}

\begin{figure}[htb]
  \centering
  \includegraphics[height=150pt,width=!]{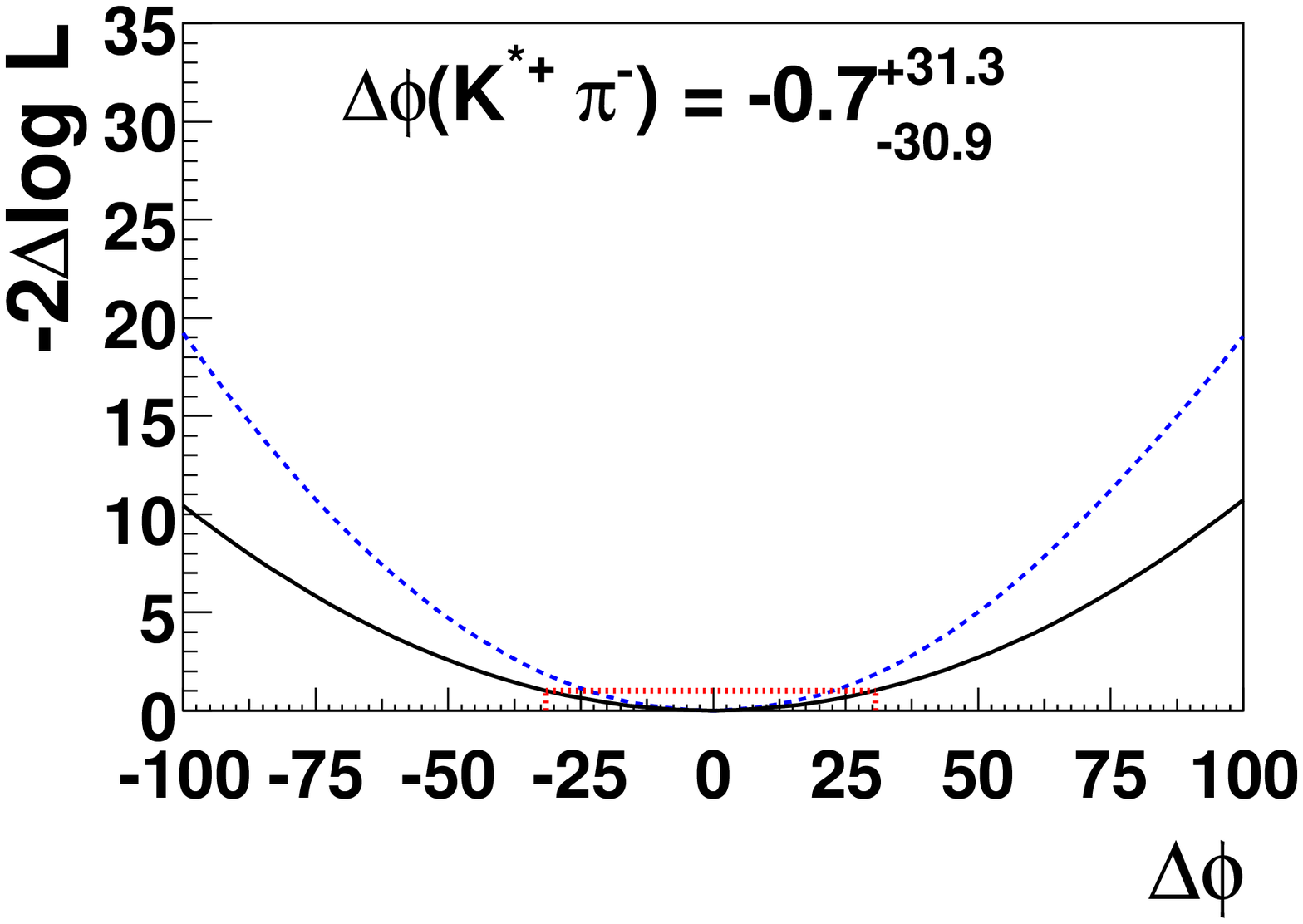}
  \includegraphics[height=150pt,width=!]{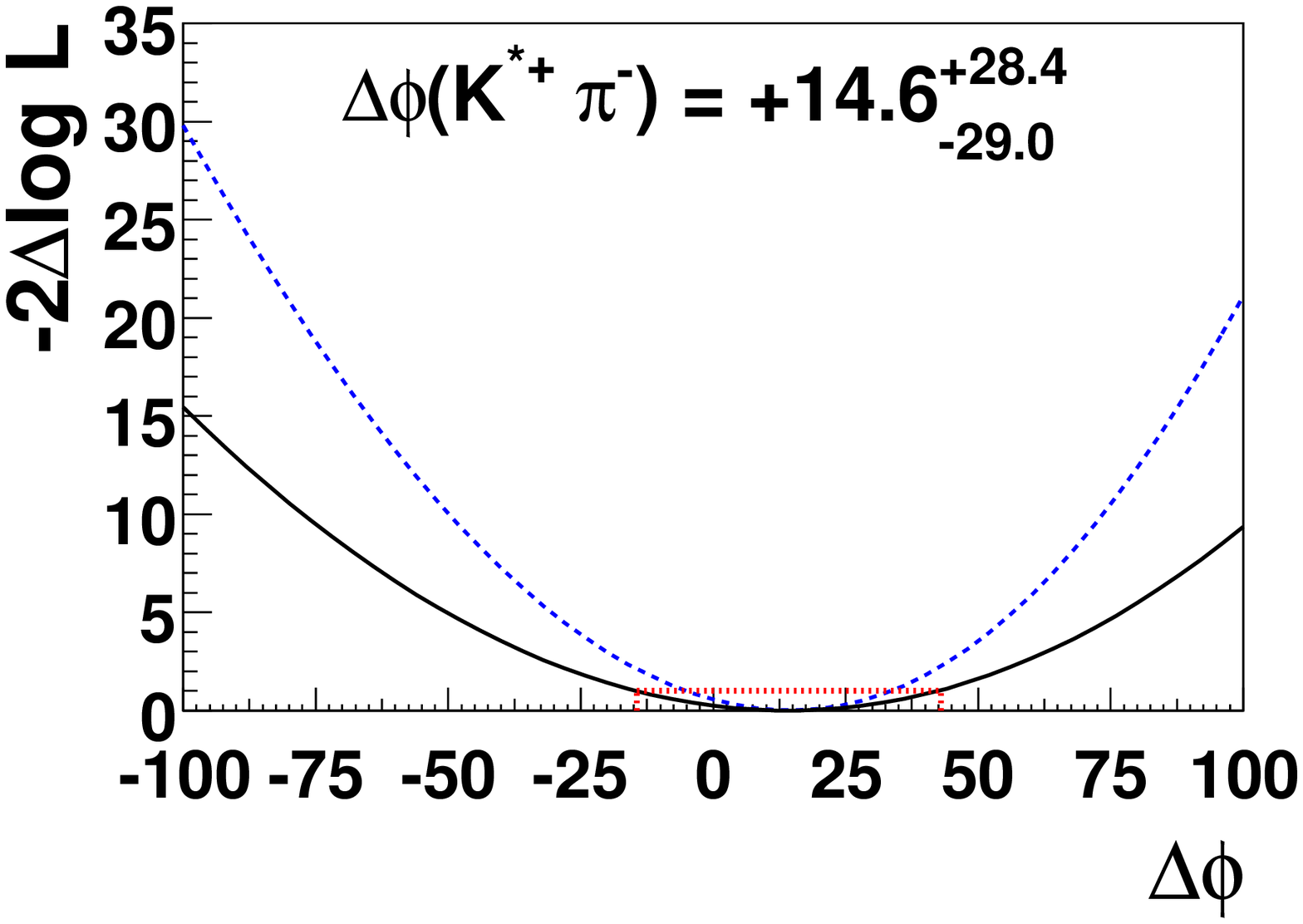}
  \caption{Likelihood scan of $\Delta \phi$ for Solution 1 (left) and Solution 2 (right). The solid (dashed) curve contains the total (statistical) error and the dotted box indicates the parameter range corresponding to $\pm 1 \sigma$.}
  \label{fig_scan4}
\end{figure}

Likelihood scans of \phione\ for both solutions are obtained by fixing \phione\ and redoing the fit. The statistical error for \Scp\ is also determined from a likelihood scan that fixes \Acp\ and \phione. Similarly, likelihood scans of $\Delta \phi$ are also produced. In addition, we also perform scans that include the systematic and model errors by convolving the likelihood with a Gaussian with width set to the quadratic sum of the systematic and model uncertainties. These are shown in Figs.~\ref{fig_scan1}-\ref{fig_scan4}.

\section{Systematic Uncertainties}
The systematic errors in the vertex reconstruction include uncertainties in the IP profile, charged track selection based on track helix errors, helix parameter corrections, \Dt\ and vertex goodness-of-fit selection, $\Delta z$ bias and SVD misalignment. The parameters for $w$ and $\Delta w$, resolution function, physics parameters, background shape and signal probability are varied by $\pm 1 \sigma$. Where histograms were used, all bins are simultaneously but independently varied by their respective errors. The effect of $CP$ asymmetry in the background is estimated by varying the $CP$ parameters of the entire neutral $B$ component within the physical range except for \etapks, which is varied within its known uncertainties. Toy MC samples showed some small fitting bias for $CP$ parameters due to low statistics in each sample. We take this bias as a systematic uncertainty. The effect of misreconstructed signal events is also investigated using signal MC. This is achieved by comparing the fit result of the signal MC sample with another fit on the same sample, which required that the events were reconstructed correctly. The efficiency histogram also includes systematic uncertainties in correction factors due to tracking, \Ks\ selection and particle identification. The systematics from these data to MC efficiency ratios are calculated from independent studies at Belle. Tag-side interference, which only affects $CP$ eigenstates, comes from $CP$ violation on the tag side~\cite{TSI}, and is estimated with $B \rightarrow D^{*} l \nu$. We generate MC pseudo-experiments and perform an ensemble test to obtain systematic biases from tag-side interference~\cite{TSI2}. In the signal model, the Dalitz plot mass, widths and other parameters are varied by their errors and the Blatt-Weisskopf barrier factors are included. The Dalitz plot model is the dominant systematic source and is quoted separately. We introduce additional resonances, $K^{*}_{2}(1430)$, $K^{*}_{0}(1680)$, $\omega(782)$, $\rho^{0}(1450)$,  $\rho^{0}(1700)$ and $f_{0}(1710)$ into the signal model to estimate possible effects from other resonances not included in the nominal model. These resonances are included separately with their amplitudes and phases as free parameters and their $CP$ parameters shared with the \ftwo, \fX\ and the non-resonant component. The amplitudes of these additional resonances are not found to be significant. The shape of the non-resonant component is empirically chosen, so different parametrizations are tested. They include modeling the non-resonant part with the tail of a Breit-Wigner, $R_{\rm NR}(s;\alpha) = i\alpha/(s+i\alpha)$ and a power law whose exponent is a fit parameter, $R_{\rm NR}(s;\alpha) = s^{-\alpha}$. The fit differences from these alternate Dalitz plot parametrizations were summed in quadrature. The systematic errors for both solutions are summarized in Table~\ref{tab_syst1} and Table~\ref{tab_syst2}. As the systematic uncertainty from the Dalitz plot model is the largest, it is quoted separately.
\begin{table}[htb]
  \caption{Summary of systematic uncertainties for Solution 1.}
  \label{tab_syst1}
  \begin{tabular}
    {@{\hspace{0.5cm}}c@{\hspace{0.5cm}}||@{\hspace{0.5cm}}c@{\hspace{0.25cm}}  @{\hspace{0.25cm}}c@{\hspace{0.25cm}}  @{\hspace{0.25cm}}c@{\hspace{0.25cm}}  @{\hspace{0.25cm}}c@{\hspace{0.25cm}}  @{\hspace{0.25cm}}c@{\hspace{0.25cm}} @{\hspace{0.25cm}}c@{\hspace{0.5cm}}}
    \hline \hline
    \begin{sideways}Category\end{sideways} & \begin{sideways}$\delta \Acp(\rho^{0} \Ks)$\end{sideways} & \begin{sideways}$\delta \phione(\rho^{0} \Ks)$\end{sideways} & \begin{sideways}$\delta \Acp(f_{0} \Ks)$\end{sideways} & \begin{sideways}$\delta \phione(f_{0} \Ks)$\end{sideways} & \begin{sideways}$\delta \Acp(\Ksp \pim)$\end{sideways} & \begin{sideways}$\delta \Delta \phi(\Ksp \pim)$\end{sideways} \\
    \hline
    Vertex Reconstruction & $0.068$ & $2.16$ & $0.045$ & $1.64$ & $0.039$ & $8.51$\\
    Flavor Tagging & $0.005$ & $0.11$ & $0.005$ & $0.20$ & $0.001$ & $0.62$\\
    \Dt\ Resolution Function & $0.022$ & $0.87$ & $0.011$ & $0.99$ & $0.013$ & $3.19$\\
    Physics Parameters & $0.000$ & $0.04$ & $0.001$ & $0.03$ & $0.001$ & $0.12$\\
    Background Model & $0.052$ & $1.52$ & $0.020$ & $1.50$ & $0.011$ & $3.36$\\
    Signal Probability & $0.019$ & $0.63$ & $0.008$ & $0.54$ & $0.007$ & $3.92$\\
    Fit Bias & $0.013$ & $1.05$ & $0.004$ & $0.07$ & $0.002$ & $0.66$\\
    Misreconstruction & $0.008$ & $0.38$ & $0.008$ & $0.34$ & $0.009$ & $0.31$\\
    Efficiency & $0.025$ & $0.33$ & $0.008$ & $0.59$ & $0.016$ & $1.54$\\
    Tag-side Interference & $0.039$ & $0.06$ & $0.043$ & $0.03$ & N/A & N/A\\
    Signal Model & $0.045$ & $0.75$ & $0.015$ & $1.02$ & $0.016$ & $2.86$\\ \hline
    {\bf Total} & $0.112$ & $3.17$ & $0.069$ & $2.79$ & $0.050$ & $10.99$\\ \hline
    {\bf Amplitude Model} & $0.097$ & $3.53$ & $0.091$ & $3.33$ & $0.053$ & $17.61$\\ \hline \hline
  \end{tabular}
\end{table}
\begin{table}[htb]
  \caption{Summary of systematic uncertainties for Solution 2.}
  \label{tab_syst2}
  \begin{tabular}
    {@{\hspace{0.5cm}}c@{\hspace{0.5cm}}||@{\hspace{0.5cm}}c@{\hspace{0.25cm}}  @{\hspace{0.25cm}}c@{\hspace{0.25cm}}  @{\hspace{0.25cm}}c@{\hspace{0.25cm}}  @{\hspace{0.25cm}}c@{\hspace{0.25cm}}  @{\hspace{0.25cm}}c@{\hspace{0.25cm}} @{\hspace{0.25cm}}c@{\hspace{0.5cm}}}
    \hline \hline
    \begin{sideways}Category\end{sideways} & \begin{sideways}$\delta \Acp(\rho^{0} \Ks)$\end{sideways} & \begin{sideways}$\delta \phione(\rho^{0} \Ks)$\end{sideways} & \begin{sideways}$\delta \Acp(f_{0} \Ks)$\end{sideways} & \begin{sideways}$\delta \phione(f_{0} \Ks)$\end{sideways} & \begin{sideways}$\delta \Acp(\Ksp \pim)$\end{sideways} & \begin{sideways}$\delta \Delta \phi(\Ksp \pim)$\end{sideways}\\
    \hline
    Vertex Reconstruction & $0.055$ & $2.47$ & $0.031$ & $1.95$ & $0.041$ & $8.87$\\
    Flavor Tagging & $0.006$ & $0.15$ & $0.005$ & $0.17$ & $0.001$ & $0.49$\\
    \Dt\ Resolution Function & $0.035$ & $0.74$ & $0.009$ & $1.00$ & $0.005$ & $3.16$\\
    Physics Parameters & $0.000$ & $0.04$ & $0.002$ & $0.03$ & $0.000$ & $0.05$\\
    Background Model & $0.078$ & $1.60$ & $0.020$ & $1.17$ & $0.014$ & $4.93$\\
    Signal Probability & $0.041$ & $0.68$ & $0.012$ & $0.42$ & $0.005$ & $1.71$\\
    Fit Bias & $0.013$ & $1.01$ & $0.004$ & $0.07$ & $0.002$ & $0.66$\\
    Misreconstruction & $0.010$ & $0.35$ & $0.010$ & $0.20$ & $0.007$ & $0.19$\\
    Efficiency & $0.017$ & $0.08$ & $0.010$ & $0.44$ & $0.028$ & $0.96$\\
    Tag-side Interference & $0.039$ & $0.06$ & $0.043$ & $0.03$ & N/A & N/A\\
    Signal Model & $0.034$ & $0.53$ & $0.020$ & $0.83$ & $0.010$ & $2.10$\\ \hline
    {\bf Total} & $0.124$ & $3.33$ & $0.064$ & $2.70$ & $0.054$ & $11.04$\\ \hline
    {\bf Amplitude Model} & $0.097$ & $3.53$ & $0.091$ & $3.33$ & $0.053$ & $17.61$\\ \hline \hline
  \end{tabular}
\end{table}

\section{Conclusion}
In summary, we perform a time-dependent Dalitz plot measurement of $CP$ parameters in \Kspipi\ decays and find two solutions that describe the data well. The first of these solutions may be preferred by external information from other measurements; however, we retain both solutions. This is Belle's first measurement of $CP$ violation parameters in the \rhozks\ channel and the first measurement of the $CP$ parameters in \fzks\ decays using a time-dependent Dalitz plot technique. There is currently no evidence for direct $CP$ violation in \rhozks, \fzks\ and \Ksppim, while mixing-induced $CP$ violation in \rhozks\ and \fzks\ decays deviates from zero by roughly $2 \sigma$ and is consistent with measurements in $b \rightarrow c \bar c s$ transitions. We also measured the phase difference between $\Bz \rightarrow \Ksp \pim$ and $\Bzb \rightarrow \Ksm \pip$, which may be used to extract $\phi_{3}$.

\section{Acknowledgments}


%
We thank the KEKB group for the excellent operation of the
accelerator, the KEK cryogenics group for the efficient
operation of the solenoid, and the KEK computer group and
the National Institute of Informatics for valuable computing
and SINET3 network support.  We acknowledge support from
the Ministry of Education, Culture, Sports, Science, and
Technology (MEXT) of Japan, the Japan Society for the 
Promotion of Science (JSPS), and the Tau-Lepton Physics 
Research Center of Nagoya University; 
the Australian Research Council and the Australian 
Department of Industry, Innovation, Science and Research;
the National Natural Science Foundation of China under
contract No.~10575109, 10775142, 10875115 and 10825524; 
the Department of Science and Technology of India; 
the BK21 program of the Ministry of Education of Korea, 
the CHEP src program and Basic Research program (grant 
No. R01-2008-000-10477-0) of the 
Korea Science and Engineering Foundation;
the Polish Ministry of Science and Higher Education;
the Ministry of Education and Science of the Russian
Federation and the Russian Federal Agency for Atomic Energy;
the Slovenian Research Agency;  the Swiss
National Science Foundation; the National Science Council
and the Ministry of Education of Taiwan; and the U.S.\
Department of Energy.
This work is supported by a Grant-in-Aid from MEXT for 
Science Research in a Priority Area ("New Development of 
Flavor Physics"), and from JSPS for Creative Scientific 
Research ("Evolution of Tau-lepton Physics").

\appendix
\begin{table}
  \caption{Statistical correlation matrix for Solution 1.}
  \label{tab_corr11}
  \small
  \begin{tabular}
    {@{\hspace{0.5cm}}c@{\hspace{0.5cm}}| @{\hspace{0.5cm}}c@{\hspace{0.25cm}}  @{\hspace{0.25cm}}c@{\hspace{0.25cm}}  @{\hspace{0.25cm}}c@{\hspace{0.25cm}}  @{\hspace{0.25cm}}c@{\hspace{0.25cm}}  @{\hspace{0.25cm}}c@{\hspace{0.25cm}}  @{\hspace{0.25cm}}c@{\hspace{0.25cm}}  @{\hspace{0.25cm}}c@{\hspace{0.25cm}}  @{\hspace{0.25cm}}c@{\hspace{0.25cm}}  @{\hspace{0.25cm}}c@{\hspace{0.25cm}}}
    \hline \hline
                     & \begin{sideways}$c_{\Ksp}$\end{sideways} & \begin{sideways}$d_{\Ksp}$\end{sideways} & \begin{sideways}$a_{\Kstarp}$\end{sideways} & \begin{sideways}$b_{\Kstarp}$\end{sideways} & \begin{sideways}$c_{\Kstarp}$\end{sideways} & \begin{sideways}$d_{\Kstarp}$\end{sideways} & \begin{sideways}$a_{\rhoz}$\end{sideways} & \begin{sideways}$b_{\rhoz}$\end{sideways} & \begin{sideways}$c_{\rhoz}$\end{sideways} \\
    \hline
    $c_{\Ksp}$                 & $+1.00$ \\
    $d_{\Ksp}$                 & $-0.06$ & $+1.00$ \\
    $a_{\Kstarp}$              & $+0.14$ & $-0.07$ & $+1.00$ \\
    $b_{\Kstarp}$              & $+0.24$ & $-0.07$ & $+0.24$ & $+1.00$ \\
    $c_{\Kstarp}$              & $-0.03$ & $+0.11$ & $+0.04$ & $+0.01$ &
    $+1.00$ \\
    $d_{\Kstarp}$              & $-0.06$ & $+0.82$ & $-0.15$ & $-0.26$ &
    $+0.15$ & $+1.00$ \\
    $a_{\rhoz}$                & $+0.08$ & $-0.06$ & $+0.47$ & $+0.10$ &
    $-0.01$ & $-0.10$ & $+1.00$ \\
    $b_{\rhoz}$                & $+0.11$ & $-0.14$ & $+0.09$ & $+0.39$ &
    $-0.04$ & $-0.24$ & $-0.04$ & $+1.00$ \\
    $c_{\rhoz}$                & $+0.07$ & $-0.09$ & $+0.04$ & $+0.11$ &
    $+0.06$ & $-0.08$ & $+0.06$ & $-0.02$ & $+1.00$ \\
    $d_{\rhoz}$                & $+0.01$ & $+0.15$ & $-0.03$ & $-0.02$ &
    $+0.09$ & $+0.22$ & $-0.07$ & $+0.07$ & $-0.01$ \\
    $a_{\fz}$                  & $+0.02$ & $+0.16$ & $+0.50$ & $-0.07$ &
    $+0.03$ & $+0.15$ & $+0.36$ & $-0.09$ & $-0.05$ \\
    $b_{\fz}$                  & $+0.18$ & $-0.15$ & $+0.18$ & $+0.53$ &
    $-0.07$ & $-0.27$ & $-0.03$ & $+0.63$ & $+0.05$ \\
    $c_{\fz}$                  & $+0.00$ & $+0.03$ & $-0.03$ & $-0.03$ &
    $-0.15$ & $-0.01$ & $+0.01$ & $+0.05$ & $-0.29$ \\
    $d_{\fz}$                  & $-0.07$ & $+0.19$ & $-0.07$ & $-0.06$ &
    $+0.01$ & $+0.22$ & $-0.01$ & $-0.08$ & $-0.06$ \\
    $a_{\ftwo}$                & $-0.07$ & $+0.17$ & $+0.20$ & $-0.15$ &
    $-0.00$ & $+0.16$ & $+0.12$ & $-0.02$ & $-0.06$ \\
    $b_{\ftwo}$                & $+0.17$ & $-0.07$ & $+0.14$ & $+0.49$ &
    $-0.03$ & $-0.17$ & $+0.09$ & $+0.44$ & $+0.10$ \\
    $c_{\rm Rest}$             & $-0.20$ & $+0.10$ & $+0.06$ & $-0.07$ &
    $+0.31$ & $+0.17$ & $-0.01$ & $-0.07$ & $-0.13$ \\
    $d_{\rm Rest}$             & $-0.10$ & $+0.82$ & $-0.17$ & $-0.29$ &
    $-0.02$ & $+0.96$ & $-0.10$ & $-0.23$ & $-0.13$ \\
    $a_{\fX}$                  & $+0.00$ & $+0.08$ & $+0.27$ & $-0.00$ &
    $+0.05$ & $+0.08$ & $+0.11$ & $-0.04$ & $-0.03$ \\
    $b_{\fX}$                  & $+0.14$ & $-0.18$ & $+0.16$ & $+0.45$ &
    $-0.04$ & $-0.28$ & $+0.14$ & $+0.45$ & $+0.09$ \\
    $a_{(\Ks \pip)_{\rm NR}}$  & $+0.08$ & $-0.00$ & $+0.80$ & $+0.24$ &
    $+0.06$ & $-0.08$ & $+0.34$ & $+0.07$ & $+0.01$ \\
    $b_{(\Ks \pip)_{\rm NR}}$  & $+0.21$ & $-0.04$ & $+0.10$ & $+0.89$ &
    $-0.04$ & $-0.23$ & $+0.12$ & $+0.38$ & $+0.09$ \\
    $a_{(\Ks \pim)_{\rm NR}}$  & $-0.25$ & $+0.40$ & $-0.25$ & $-0.50$ &
    $+0.03$ & $+0.48$ & $-0.16$ & $-0.18$ & $-0.18$ \\
    $b_{(\Ks \pim)_{\rm NR}}$  & $+0.11$ & $+0.19$ & $+0.07$ & $+0.26$ &
    $+0.05$ & $+0.18$ & $-0.27$ & $+0.23$ & $-0.05$ \\
    $a_{(\pip \pim)_{\rm NR}}$ & $-0.14$ & $+0.13$ & $+0.10$ & $-0.30$ &
    $+0.02$ & $+0.13$ & $-0.13$ & $-0.05$ & $-0.07$ \\
    $b_{(\pip \pim)_{\rm NR}}$ & $+0.00$ & $+0.09$ & $-0.08$ & $+0.25$ &
    $-0.07$ & $-0.01$ & $+0.06$ & $+0.49$ & $-0.01$ \\
    $\alpha$                   & $+0.17$ & $-0.31$ & $+0.38$ & $+0.51$ &
    $-0.02$ & $-0.41$ & $+0.32$ & $+0.14$ & $+0.15$ \\
    \hline \hline
  \end{tabular}
\end{table}

\begin{table}
  \caption{Statistical correlation matrix for Solution 1.}
  \label{tab_corr12}
  \small
  \begin{tabular}
    {@{\hspace{0.5cm}}c@{\hspace{0.5cm}}| @{\hspace{0.5cm}}c@{\hspace{0.25cm}}  @{\hspace{0.25cm}}c@{\hspace{0.25cm}}  @{\hspace{0.25cm}}c@{\hspace{0.25cm}}  @{\hspace{0.25cm}}c@{\hspace{0.25cm}}  @{\hspace{0.25cm}}c@{\hspace{0.25cm}}  @{\hspace{0.25cm}}c@{\hspace{0.25cm}}  @{\hspace{0.25cm}}c@{\hspace{0.25cm}}  @{\hspace{0.25cm}}c@{\hspace{0.25cm}}  @{\hspace{0.25cm}}c@{\hspace{0.25cm}}}
    \hline \hline
                     & \begin{sideways}$d_{\rhoz}$\end{sideways} & \begin{sideways}$a_{\fz}$\end{sideways} & \begin{sideways}$b_{\fz}$\end{sideways} & \begin{sideways}$c_{\fz}$\end{sideways} & \begin{sideways}$d_{\fz}$\end{sideways} & \begin{sideways}$a_{\ftwo}$\end{sideways} & \begin{sideways}$b_{\ftwo}$\end{sideways} & \begin{sideways}$c_{\rm Rest}$\end{sideways} & \begin{sideways}$d_{\rm Rest}$\end{sideways} \\
    \hline
    $d_{\rhoz}$                & $+1.00$ \\
    $a_{\fz}$                  & $-0.03$ & $+1.00$ \\
    $b_{\fz}$                  & $-0.01$ & $-0.19$ & $+1.00$ \\
    $c_{\fz}$                  & $-0.01$ & $-0.02$ & $-0.01$ & $+1.00$ \\
    $d_{\fz}$                  & $+0.37$ & $+0.02$ & $-0.14$ & $+0.03$ &
    $+1.00$ \\
    $a_{\ftwo}$                & $-0.02$ & $+0.37$ & $-0.05$ & $+0.05$ &
    $-0.07$ & $+1.00$ \\
    $b_{\ftwo}$                & $-0.04$ & $+0.05$ & $+0.51$ & $-0.05$ &
    $-0.07$ & $-0.01$ & $+1.00$ \\
    $c_{\rm Rest}$             & $+0.09$ & $+0.09$ & $-0.12$ & $-0.25$ &
    $-0.08$ & $+0.04$ & $-0.06$ & $+1.00$ \\
    $d_{\rm Rest}$             & $+0.18$ & $+0.15$ & $-0.26$ & $+0.06$ &
    $+0.22$ & $+0.18$ & $-0.17$ & $+0.11$ & $+1.00$ \\
    $a_{\fX}$                  & $+0.04$ & $+0.18$ & $-0.09$ & $-0.00$ &
    $+0.04$ & $+0.07$ & $-0.09$ & $+0.11$ & $+0.07$ \\
    $b_{\fX}$                  & $-0.04$ & $-0.07$ & $+0.47$ & $+0.01$ &
    $-0.00$ & $-0.18$ & $+0.54$ & $-0.10$ & $-0.29$ \\
    $a_{(\Ks \pip)_{\rm NR}}$  & $-0.00$ & $+0.40$ & $+0.08$ & $+0.01$ &
    $-0.05$ & $+0.23$ & $+0.03$ & $+0.08$ & $-0.10$ \\
    $b_{(\Ks \pip)_{\rm NR}}$  & $-0.04$ & $-0.05$ & $+0.54$ & $-0.01$ &
    $-0.05$ & $-0.08$ & $+0.47$ & $-0.09$ & $-0.23$ \\
    $a_{(\Ks \pim)_{\rm NR}}$  & $+0.07$ & $+0.24$ & $-0.42$ & $+0.06$ &
    $+0.15$ & $+0.32$ & $-0.28$ & $+0.08$ & $+0.51$ \\
    $b_{(\Ks \pim)_{\rm NR}}$  & $+0.08$ & $+0.01$ & $+0.20$ & $-0.00$ &
    $+0.10$ & $-0.16$ & $+0.17$ & $+0.07$ & $+0.17$ \\
    $a_{(\pip \pim)_{\rm NR}}$ & $+0.05$ & $-0.10$ & $+0.02$ & $+0.03$ &
    $-0.04$ & $+0.28$ & $-0.19$ & $+0.07$ & $+0.15$ \\
    $b_{(\pip \pim)_{\rm NR}}$ & $-0.04$ & $-0.08$ & $+0.50$ & $+0.05$ &
    $-0.03$ & $+0.14$ & $+0.35$ & $-0.11$ & $+0.03$ \\
    $\alpha$                   & $-0.07$ & $-0.08$ & $+0.29$ & $-0.02$ &
    $-0.16$ & $-0.06$ & $+0.18$ & $-0.10$ & $-0.43$ \\
    \hline \hline
  \end{tabular}
\end{table}

\begin{table}
  \caption{Statistical correlation matrix for Solution 1.}
  \label{tab_corr13}
  \small
  \begin{tabular}
    {@{\hspace{0.5cm}}c@{\hspace{0.5cm}}| @{\hspace{0.5cm}}c@{\hspace{0.25cm}}  @{\hspace{0.25cm}}c@{\hspace{0.25cm}}  @{\hspace{0.25cm}}c@{\hspace{0.25cm}}  @{\hspace{0.25cm}}c@{\hspace{0.25cm}}  @{\hspace{0.25cm}}c@{\hspace{0.25cm}}  @{\hspace{0.25cm}}c@{\hspace{0.25cm}}  @{\hspace{0.25cm}}c@{\hspace{0.25cm}}  @{\hspace{0.25cm}}c@{\hspace{0.25cm}}  @{\hspace{0.25cm}}c@{\hspace{0.25cm}}}
    \hline \hline
                     & \begin{sideways}$a_{\fX}$\end{sideways} & \begin{sideways}$b_{\fX}$\end{sideways} & \begin{sideways}$a_{(\Ks \pip)_{\rm NR}}$\end{sideways} & \begin{sideways}$b_{(\Ks \pip)_{\rm NR}}$\end{sideways} & \begin{sideways}$a_{(\Ks \pim)_{\rm NR}}$\end{sideways} & \begin{sideways}$b_{(\Ks \pim)_{\rm NR}}$\end{sideways} & \begin{sideways}$a_{(\pip \pim)_{\rm NR}}$\end{sideways} & \begin{sideways}$b_{(\pip \pim)_{\rm NR}}$\end{sideways} & \begin{sideways}$\alpha$\end{sideways} \\
    \hline
    $a_{\fX}$                  & $+1.00$ \\
    $b_{\fX}$                  & $+0.13$ & $+1.00$ \\
    $a_{(\Ks \pip)_{\rm NR}}$  & $+0.28$ & $+0.10$ & $+1.00$ \\
    $b_{(\Ks \pip)_{\rm NR}}$  & $-0.05$ & $+0.44$ & $+0.13$ & $+1.00$ \\
    $a_{(\Ks \pim)_{\rm NR}}$  & $+0.13$ & $-0.34$ & $-0.10$ & $-0.47$ &
    $+1.00$ \\
    $b_{(\Ks \pim)_{\rm NR}}$  & $+0.12$ & $+0.10$ & $+0.09$ & $+0.24$ &
    $-0.02$ & $+1.00$ \\
    $a_{(\pip \pim)_{\rm NR}}$ & $+0.17$ & $-0.10$ & $+0.28$ & $-0.21$ &
    $+0.31$ & $-0.07$ & $+1.00$ \\
    $b_{(\pip \pim)_{\rm NR}}$ & $-0.16$ & $+0.46$ & $-0.07$ & $+0.44$ &
    $+0.18$ & $+0.06$ & $+0.11$ & $+1.00$ \\
    $\alpha$                   & $-0.11$ & $+0.28$ & $+0.42$ & $+0.35$ &
    $-0.57$ & $-0.35$ & $-0.34$ & $-0.01$ & $+1.00$ \\
    \hline \hline
  \end{tabular}
\end{table}

\begin{table}
  \caption{Statistical correlation matrix for Solution 2.}
  \label{tab_corr21}
  \small
  \begin{tabular}
    {@{\hspace{0.5cm}}c@{\hspace{0.5cm}}| @{\hspace{0.5cm}}c@{\hspace{0.25cm}}  @{\hspace{0.25cm}}c@{\hspace{0.25cm}}  @{\hspace{0.25cm}}c@{\hspace{0.25cm}}  @{\hspace{0.25cm}}c@{\hspace{0.25cm}}  @{\hspace{0.25cm}}c@{\hspace{0.25cm}}  @{\hspace{0.25cm}}c@{\hspace{0.25cm}}  @{\hspace{0.25cm}}c@{\hspace{0.25cm}}  @{\hspace{0.25cm}}c@{\hspace{0.25cm}}  @{\hspace{0.25cm}}c@{\hspace{0.25cm}}}
    \hline \hline
                     & \begin{sideways}$c_{\Ksp}$\end{sideways} & \begin{sideways}$d_{\Ksp}$\end{sideways} & \begin{sideways}$a_{\Kstarp}$\end{sideways} & \begin{sideways}$b_{\Kstarp}$\end{sideways} & \begin{sideways}$c_{\Kstarp}$\end{sideways} & \begin{sideways}$d_{\Kstarp}$\end{sideways} & \begin{sideways}$a_{\rhoz}$\end{sideways} & \begin{sideways}$b_{\rhoz}$\end{sideways} & \begin{sideways}$c_{\rhoz}$\end{sideways} \\
    \hline
    $c_{\Ksp}$                 & $+1.00$ \\
    $d_{\Ksp}$                 & $+0.05$ & $+1.00$ \\
    $a_{\Kstarp}$              & $+0.05$ & $+0.18$ & $+1.00$ \\
    $b_{\Kstarp}$              & $+0.17$ & $+0.28$ & $+0.35$ & $+1.00$ \\
    $c_{\Kstarp}$              & $-0.01$ & $+0.05$ & $+0.18$ & $+0.06$ & $+1.00$ \\
    $d_{\Kstarp}$              & $+0.02$ & $+0.61$ & $+0.23$ & $+0.05$ & $+0.38$ & $+1.00$ \\
    $a_{\rhoz}$                & $+0.04$ & $+0.13$ & $-0.02$ & $-0.08$ & $-0.10$ & $+0.03$ & $+1.00$ \\
    $b_{\rhoz}$                & $+0.03$ & $-0.09$ & $+0.20$ & $+0.27$ & $+0.03$ & $-0.13$ & $-0.58$ & $+1.00$ \\
    $c_{\rhoz}$                & $+0.02$ & $+0.16$ & $-0.06$ & $-0.00$ & $-0.01$ & $+0.13$ & $+0.23$ & $-0.42$ & $+1.00$ \\
    $d_{\rhoz}$                & $+0.00$ & $+0.25$ & $+0.05$ & $+0.07$ & $+0.12$ & $+0.32$ & $-0.06$ & $+0.05$ & $-0.12$ \\
    $a_{\fz}$                  & $+0.08$ & $+0.12$ & $+0.29$ & $+0.03$ & $-0.03$ & $+0.03$ & $+0.49$ & $-0.23$ & $+0.15$ \\
    $b_{\fz}$                  & $+0.05$ & $-0.03$ & $+0.19$ & $+0.35$ & $-0.02$ & $-0.13$ & $-0.46$ & $+0.73$ & $-0.30$ \\
    $c_{\fz}$                  & $+0.01$ & $-0.09$ & $+0.00$ & $-0.02$ & $-0.04$ & $-0.11$ & $-0.07$ & $+0.19$ & $-0.40$ \\
    $d_{\fz}$                  & $-0.05$ & $+0.22$ & $+0.06$ & $+0.04$ & $+0.03$ & $+0.27$ & $+0.02$ & $+0.00$ & $-0.13$ \\
    $a_{\ftwo}$                & $+0.00$ & $+0.03$ & $+0.08$ & $-0.05$ & $-0.03$ & $-0.04$ & $+0.02$ & $+0.19$ & $-0.06$ \\
    $b_{\ftwo}$                & $+0.11$ & $+0.10$ & $+0.08$ & $+0.40$ & $-0.07$ & $-0.06$ & $-0.00$ & $+0.28$ & $+0.04$ \\
    $c_{\rm Rest}$             & $-0.16$ & $+0.18$ & $-0.03$ & $-0.12$ & $-0.21$ & $+0.28$ & $+0.15$ & $-0.13$ & $-0.02$ \\
    $d_{\rm Rest}$             & $+0.02$ & $+0.66$ & $+0.03$ & $+0.01$ & $-0.21$ & $+0.68$ & $+0.11$ & $-0.15$ & $+0.18$ \\
    $a_{\fX}$                  & $+0.04$ & $+0.08$ & $+0.09$ & $-0.04$ & $-0.04$ & $+0.02$ & $+0.29$ & $-0.06$ & $+0.01$ \\
    $b_{\fX}$                  & $+0.05$ & $+0.06$ & $-0.09$ & $+0.25$ & $-0.08$ & $-0.09$ & $+0.05$ & $+0.25$ & $+0.00$ \\
    $a_{(\Ks \pip)_{\rm NR}}$  & $+0.01$ & $+0.02$ & $+0.03$ & $-0.04$ & $-0.04$ & $-0.13$ & $+0.48$ & $-0.14$ & $+0.08$ \\
    $b_{(\Ks \pip)_{\rm NR}}$  & $+0.16$ & $+0.14$ & $-0.36$ & $+0.51$ & $-0.20$ & $-0.19$ & $+0.26$ & $-0.01$ & $+0.10$ \\
    $a_{(\Ks \pim)_{\rm NR}}$  & $-0.01$ & $+0.11$ & $+0.39$ & $+0.20$ & $+0.10$ & $+0.13$ & $-0.20$ & $+0.26$ & $-0.10$ \\
    $b_{(\Ks \pim)_{\rm NR}}$  & $+0.15$ & $+0.12$ & $-0.04$ & $+0.38$ & $-0.13$ & $-0.05$ & $+0.37$ & $-0.36$ & $+0.27$ \\
    $a_{(\pip \pim)_{\rm NR}}$ & $-0.07$ & $-0.10$ & $+0.16$ & $-0.10$ & $+0.06$ & $-0.09$ & $-0.29$ & $+0.51$ & $-0.23$ \\
    $b_{(\pip \pim)_{\rm NR}}$ & $+0.02$ & $+0.11$ & $+0.07$ & $+0.34$ & $-0.10$ & $-0.06$ & $+0.09$ & $+0.18$ & $+0.09$ \\
    $\alpha$                   & $-0.01$ & $-0.05$ & $-0.42$ & $-0.11$ & $-0.04$ & $-0.13$ & $-0.05$ & $+0.13$ & $-0.12$ \\
    \hline \hline
  \end{tabular}
\end{table}

\begin{table}
  \caption{Statistical correlation matrix for Solution 2.}
  \label{tab_corr22}
  \small
  \begin{tabular}
    {@{\hspace{0.5cm}}c@{\hspace{0.5cm}}| @{\hspace{0.5cm}}c@{\hspace{0.25cm}}  @{\hspace{0.25cm}}c@{\hspace{0.25cm}}  @{\hspace{0.25cm}}c@{\hspace{0.25cm}}  @{\hspace{0.25cm}}c@{\hspace{0.25cm}}  @{\hspace{0.25cm}}c@{\hspace{0.25cm}}  @{\hspace{0.25cm}}c@{\hspace{0.25cm}}  @{\hspace{0.25cm}}c@{\hspace{0.25cm}}  @{\hspace{0.25cm}}c@{\hspace{0.25cm}}  @{\hspace{0.25cm}}c@{\hspace{0.25cm}}}
    \hline \hline
                     & \begin{sideways}$d_{\rhoz}$\end{sideways} & \begin{sideways}$a_{\fz}$\end{sideways} & \begin{sideways}$b_{\fz}$\end{sideways} & \begin{sideways}$c_{\fz}$\end{sideways} & \begin{sideways}$d_{\fz}$\end{sideways} & \begin{sideways}$a_{\ftwo}$\end{sideways} & \begin{sideways}$b_{\ftwo}$\end{sideways} & \begin{sideways}$c_{\rm Rest}$\end{sideways} & \begin{sideways}$d_{\rm Rest}$\end{sideways} \\
    \hline
    $d_{\rhoz}$                & $+1.00$ \\
    $a_{\fz}$                  & $-0.07$ & $+1.00$ \\
    $b_{\fz}$                  & $+0.02$ & $-0.17$ & $+1.00$ \\
    $c_{\fz}$                  & $-0.05$ & $-0.08$ & $+0.11$ & $+1.00$ \\
    $d_{\fz}$                  & $+0.43$ & $-0.03$ & $+0.05$ & $+0.14$ & $+1.00$ \\
    $a_{\ftwo}$                & $-0.01$ & $+0.32$ & $+0.24$ & $+0.01$ & $-0.09$ & $+1.00$ \\
    $b_{\ftwo}$                & $+0.01$ & $+0.17$ & $+0.39$ & $-0.07$ & $+0.04$ & $+0.15$ & $+1.00$ \\
    $c_{\rm Rest}$             & $-0.15$ & $+0.11$ & $-0.08$ & $-0.06$ & $-0.05$ & $+0.03$ & $+0.00$ & $+1.00$ \\
    $d_{\rm Rest}$             & $+0.33$ & $+0.04$ & $-0.11$ & $-0.05$ & $+0.33$ & $+0.01$ & $-0.01$ & $+0.24$ & $+1.00$ \\
    $a_{\fX}$                  & $+0.02$ & $+0.32$ & $-0.15$ & $+0.03$ & $-0.02$ & $+0.24$ & $+0.00$ & $+0.09$ & $+0.06$ \\
    $b_{\fX}$                  & $-0.00$ & $+0.12$ & $+0.29$ & $+0.02$ & $+0.10$ & $+0.05$ & $+0.58$ & $+0.03$ & $-0.02$ \\
    $a_{(\Ks \pip)_{\rm NR}}$  & $-0.07$ & $+0.50$ & $-0.05$ & $-0.01$ & $-0.03$ & $+0.39$ & $+0.03$ & $+0.07$ & $-0.05$ \\
    $b_{(\Ks \pip)_{\rm NR}}$  & $-0.03$ & $+0.13$ & $+0.13$ & $-0.04$ & $-0.05$ & $+0.05$ & $+0.37$ & $-0.02$ & $-0.03$ \\
    $a_{(\Ks \pim)_{\rm NR}}$  & $+0.02$ & $+0.14$ & $+0.27$ & $+0.01$ & $+0.08$ & $+0.13$ & $+0.05$ & $+0.01$ & $+0.05$ \\
    $b_{(\Ks \pim)_{\rm NR}}$  & $-0.06$ & $+0.20$ & $-0.09$ & $-0.14$ & $+0.02$ & $-0.22$ & $+0.21$ & $+0.04$ & $+0.02$ \\
    $a_{(\pip \pim)_{\rm NR}}$ & $+0.03$ & $-0.25$ & $+0.52$ & $+0.09$ & $-0.01$ & $+0.25$ & $+0.07$ & $-0.06$ & $-0.10$ \\
    $b_{(\pip \pim)_{\rm NR}}$ & $-0.12$ & $+0.04$ & $+0.37$ & $-0.01$ & $+0.01$ & $-0.04$ & $+0.44$ & $+0.04$ & $-0.01$ \\
    $\alpha$                   & $+0.01$ & $+0.01$ & $+0.03$ & $+0.06$ & $-0.07$ & $+0.33$ & $-0.06$ & $-0.01$ & $-0.03$ \\
    \hline \hline
  \end{tabular}
\end{table}

\begin{table}
  \caption{Statistical correlation matrix for Solution 2.}
  \label{tab_corr23}
  \small
  \begin{tabular}
    {@{\hspace{0.5cm}}c@{\hspace{0.5cm}}| @{\hspace{0.5cm}}c@{\hspace{0.25cm}}  @{\hspace{0.25cm}}c@{\hspace{0.25cm}}  @{\hspace{0.25cm}}c@{\hspace{0.25cm}}  @{\hspace{0.25cm}}c@{\hspace{0.25cm}}  @{\hspace{0.25cm}}c@{\hspace{0.25cm}}  @{\hspace{0.25cm}}c@{\hspace{0.25cm}}  @{\hspace{0.25cm}}c@{\hspace{0.25cm}}  @{\hspace{0.25cm}}c@{\hspace{0.25cm}}  @{\hspace{0.25cm}}c@{\hspace{0.25cm}}}
    \hline \hline
                     & \begin{sideways}$a_{\fX}$\end{sideways} & \begin{sideways}$b_{\fX}$\end{sideways} & \begin{sideways}$a_{(\Ks \pip)_{\rm NR}}$\end{sideways} & \begin{sideways}$b_{(\Ks \pip)_{\rm NR}}$\end{sideways} & \begin{sideways}$a_{(\Ks \pim)_{\rm NR}}$\end{sideways} & \begin{sideways}$b_{(\Ks \pim)_{\rm NR}}$\end{sideways} & \begin{sideways}$a_{(\pip \pim)_{\rm NR}}$\end{sideways} & \begin{sideways}$b_{(\pip \pim)_{\rm NR}}$\end{sideways} & \begin{sideways}$\alpha$\end{sideways} \\
    \hline
    $a_{\fX}$                  & $+1.00$ \\
    $b_{\fX}$                  & $+0.27$ & $+1.00$ \\
    $a_{(\Ks \pip)_{\rm NR}}$  & $+0.42$ & $+0.17$ & $+1.00$ \\
    $b_{(\Ks \pip)_{\rm NR}}$  & $+0.09$ & $+0.34$ & $+0.13$ & $+1.00$ \\
    $a_{(\Ks \pim)_{\rm NR}}$  & $+0.03$ & $-0.08$ & $-0.01$ & $-0.24$ & $+1.00$ \\
    $b_{(\Ks \pim)_{\rm NR}}$  & $-0.10$ & $+0.05$ & $+0.04$ & $+0.57$ & $-0.07$ & $+1.00$ \\
    $a_{(\pip \pim)_{\rm NR}}$ & $+0.09$ & $+0.26$ & $+0.30$ & $-0.30$ & $-0.04$ & $-0.49$ & $+1.00$ \\
    $b_{(\pip \pim)_{\rm NR}}$ & $-0.25$ & $+0.44$ & $+0.03$ & $+0.34$ & $+0.10$ & $+0.34$ & $+0.15$ & $+1.00$ \\
    $\alpha$                   & $+0.31$ & $+0.13$ & $+0.37$ & $+0.22$ & $+0.14$ & $-0.36$ & $+0.05$ & $-0.37$ & $+1.00$ \\
    \hline \hline
  \end{tabular}
\end{table}

\end{document}